\documentclass[twocolumn, astrosymb]{aastex631}

\usepackage{apjfonts}
\usepackage{amssymb, amsmath}
\usepackage{graphicx}
\usepackage{CJKutf8}
\usepackage{natbib}
\usepackage{float}
\usepackage{hyperref}
\usepackage{comment}

\DeclareGraphicsExtensions{.pdf,.png,.jpg}

\def\apjs{{ApJS}}

\def\h{\hskip -3 mm}

\def\ser{{S\'{e}rsic}}

\def\logm10{{$log_{10} (M_{\star,R>10kpc}/M_{\odot})$}}
\def\logm20{{$log_{10} (M_{\star,R>20kpc}/M_{\odot})$}}

\def\mh200c{{$M_{\mathrm{200c}}$}}

\definecolor{LightGray}{gray}{0.85}
\definecolor{Tab1}{RGB}{114, 158, 206}
\definecolor{Tab2}{RGB}{255, 158,  74}
\definecolor{Tab3}{RGB}{103, 191,  92}
\definecolor{Tab4}{RGB}{174, 199, 232}
\definecolor{Tab5}{RGB}{255, 187, 120}
\definecolor{Tab6}{RGB}{152, 223, 138}
\definecolor{Tab7}{RGB}{255, 152, 150}
\definecolor{Tab8}{RGB}{197, 176, 213}
\definecolor{hpurple}{HTML}{7E16DF}

\submitjournal{AAS Journal}

\shorttitle{Zangetsu: an Isolated, Quiescent and Distorted UDG}
\shortauthors{Wei et al.}

\begin{document}

\begin{CJK*}{UTF8}{gbsn}


\title{Zangetsu: A Candidate Isolated, Quiescent, and Backsplash Ultra-Diffuse Galaxy in the COSMOS Field}

\correspondingauthor{LeYao Wei (韦乐瑶)}
    \email{weily23@mails.tsinghua.edu.cn, shuang@tsinghua.edu.cn}

\author[0009-0003-4545-0945]{Le-Yao Wei (韦乐瑶)}
    \affiliation{Department of Astronomy, Tsinghua University, Beijing 100084, China}

\author[0000-0003-1385-7591]{Song Huang (黄崧)}
    \affiliation{Department of Astronomy, Tsinghua University, Beijing 100084, China}

\author[0000-0001-9592-4190]{Jiaxuan Li (李嘉轩)}
    \affiliation{Department of Astrophysical Sciences, 4 Ivy Lane, Princeton University, Princeton, NJ 08544, USA}

\author[0000-0002-8246-7792]{Zechang Sun (孙泽昌)}
    \affiliation{Department of Astronomy, Tsinghua University, Beijing 100084, China}

\author[0000-0001-6251-649X]{Mingyu Li (李明宇)}
    \affiliation{Department of Astronomy, Tsinghua University, Beijing 100084, China}

\author[0000-0002-7563-7618]{Jiaxin Tang (唐佳欣)}
    \affiliation{Department of Astronomy, Tsinghua University, Beijing 100084, China}


\begin{abstract}

    Deep imaging surveys have changed our view of the low surface brightness (LSB) Universe. The ``renaissance'' of the LSB galaxy population, as a prime example of this recent development, continues to challenge our understanding of galaxy formation. Here, we report the serendipitous discovery of Zangetsu, an isolated, quiescent, and distorted ultra-diffuse galaxy (UDG) candidate in the COSMOS field, using images from the Hyper Suprime-Cam Subaru Strategic Program (HSC-SSP). Zangetsu exhibits an extremely low central surface brightness ($\mathrm{\mu_{0,g}}=26.60\pm0.01$ mag arcsec$^{-2}$), a very shallow inner surface brightness profile ($\mathrm{n}_{\rm Sersic}=0.40\pm0.01$), and a large angular size ($\mathrm{R_e}\approx 10.44$ arcsec). Surprisingly, Zangetsu also has a quiescent stellar population ($\mathrm{g-i}=0.96$), an unusually elongated shape ($\mathrm{b/a}\sim 0.25$), and mild morphological asymmetry, making it a rare case among known UDGs. Surface brightness fluctuation analysis of HSC and Hubble Space Telescope (HST) images only provides a distance lower limit of $D>25.4$ Mpc (thus $\mathrm{R_e}>1.38$ kpc). However, Zangetsu remains an extreme outlier in the luminosity-size relation of known LSB galaxies, suggesting that it could be an exceptionally large and/or diffuse system. Classic internal or external UDG formation mechanisms alone struggle to explain such a system. A backsplash origin may account for its isolation and quiescent nature. This finding also raises the possibility that current works may overlook similarly extreme, elongated systems that could further our understanding of the LSB Universe.
    
\end{abstract}

\keywords{
    Dwarf galaxies (416) ---
    Galaxy photometry (611) ---
    Galaxy structure (622) ---
    Low surface brightness galaxies (940)
}

\section{Introduction}
    \label{sec:intro}

    Almost a half-century ago, \citet{1976NatureDisney_LSB} predicted that, below the surface brightness limit accessible at the time, a whole new Universe awaits our discovery. We would have a biased view of galaxy formation and evolution if we could not dig deeper into the background of the sky. Such a bold prophecy has been gradually fulfilled in the following years (e.g., \citealt{Sandage1984LSB, Bothun1987LSB, Impey1988LSB, McGaugh1995LSB, Dalcanton1997LSB}). Today, the study of the low surface brightness (LSB) Universe has officially been revived by the arrival of increasingly deeper imaging data. It has become an indispensable frontier for understanding galaxies. Among many relevant progress in the last decade, the discovery of the Ultra-Diffuse Galaxies (UDGs) population is a prime example of the ongoing ``renaissance'' of the study of the LSB Universe (e.g., \citealt{van2015forty, koda2015coma1000, mihos2015virgo_galaxies, di2017fieldudgSF, leisman2017alfalfaUDG, Van2018DMdefficient, van2017GC_rich, Greco2018LSB, jiang2019fieldudgSF, carleton2019formation, benavides2021backsplash, Tanoglidis2021LSB, van2022trailHighV, Zaritsky2023SMUDGesV, Iodice2023LEWIS, Mancera2024AGC114905}, and many more). In particular, the surprising identification of 47 ``fragile'' UDGs in the central region of the Coma cluster by \citet{van2015forty} challenges our understanding of galaxy formation at the lower mass end. It ignites an ongoing debate about its nature. Following this pivotal work, not only have many more UDG (candidates) been identified in different clusters (e.g., \citealt{koda2015coma1000, Wittmann2017Perseus,lim2020virgo, marleau2024euclid, venhola2017fornax, iodice2020hydra}), they have been spotted in galaxy groups (e.g., \citealt{Roman2017groupUDG, spekkens2018atomicGroupUDG, li2023UPG}) and even in isolation (or the ``field''; e.g., \citealt{leisman2017alfalfaUDG, marleau2021ultraLowDensity}). 
    
    Recently, the Systematically Measuring Ultra-Diffuse Galaxies (SMUDGes; e.g., \citealt{Zaritsky2019SMUDGesI, Zaritsky2021SMUDGesII, Zaritsky2022SMUDGesIII, Zaritsky2023SMUDGesIV, Zaritsky2023SMUDGesV, Zaritsky2024SMUDGesVI}) project has identified $\sim7000$ UDG candidates across the whole dynamical range of environments. The statistically significant sample size provided by these recent works allows us to discuss their role in the entire LSB galaxy family as a unique population (e.g., \citealt{lim2020virgo}), debate about their dark matter content (e.g., \citealt{Van2018DMdefficient,2019DMdominatedDF44}), and characterize the globular cluster distributions in UDGs (e.g., \citealt{van2017GC_rich, Danieli2022, 2025MNRASBuzzo_GCnum}). At the same time, deeper spectroscopic follow-ups and multi-wavelength observations have started to uncover their diverse stellar population and neutral gas properties (e.g., \citealt{Gu2018StellarPop, Ferre2018StellarPop, Ruiz2018StellarPop, leisman2017alfalfaUDG, Papastergis2017HIUDG, Mancera2020HIUDG}). In general, we find UDGs everywhere we look. They could be old or young, extremely gas-rich or completely dry, decorated with many globular clusters or almost none. These discoveries are inspiring many theoretical attempts to explain such diverse physical properties organically, invoking different internal (e.g., ``failed galaxies'': \citealt{van2015forty}; high-spin halo: \citealt{2016MNRASAmorisco_highspin, 2017MNRASROng_highspin, van2022UDGspinhalo, benavides2024spinhalo}; or feedback from starburst: \citealt{di2017fieldudgSF}; and more) or external (e.g., high-velocity or early galaxy mergers: \citealt{van2022trailHighV, Wright2021}); collisional debris of past interaction: \citealt{ploeckinger2018tidal}; gravitational interaction: \citealt{poggianti2019ramstripping, conselice2018ultra}; and other environmental effects: \citealt{jiang2019fieldudgSF, tremmel2020UDGformationClu}) mechanisms to shape galaxies into the form of a UDG. Yet, despite these efforts, challenges remain. 

    Among these challenges, quenched UDGs in an isolated environment have emerged as an exciting new puzzle (e.g., DGSAT~I; \citealp{2022DGSATI}). This is not only because most field UDGs are blue and star-forming (e.g., \citealt{leisman2017alfalfaUDG, Prole2019fieldUDG}). More importantly, isolated quiescent dwarfs are statistically rare, comprising $<0.06\%$ of the dwarf galaxy population (e.g., \citealt{geha2012stellar}), regardless of whether they are diffuse enough to be UDGs. So far, the confirmed cases of quenched isolated dwarfs are small  \footnote{A list of quenched isolated dwarf galaxies: \url{https://avapolzin.github.io/projects/quench_list/}} (e.g., \citealt{sand2022tucanaB, casey2023Blobby, kado2024UGC5205, li2024hedgehog, Sand2024Sculptor}), with only three potential UDGs among them (e.g., \citealt{2022DGSATI, shen2024first}). However, new theories are being developed to explain them. For instance, \citet{2013dwarfcosmicweb} and \citet{benavides2025environmental} proposed that the ram pressure stripping from the cosmic web could quench isolated dwarfs. Meanwhile, the ``backsplash'' mechanism, where a galaxy can be ``ejected'' from a massive dark matter halo, typically losing a significant fraction of its gas while its stellar component is also disturbed, could be another viable solution for creating isolated quenched dwarf galaxies (e.g., \citealt{Balogh2000backsplash, Teyssier2012backsplash}). Well known by the simulation community (e.g., \citealp{Wetzel2014backsplash_sim, More2015backsplash_sim, Buck2019backsplash_sim, benavides2025environmental, Bhattacharyya2025}), this unique mechanism has explicitly been proposed to explain the quiescent UDGs in the field (e.g., \citealt{benavides2021backsplash}). In \citet{benavides2021backsplash}, the authors showed that backsplashed UDGs can be found a few Mpc (or $>2 \times R_{200}$) away from the previous host halos in their simulations and predicted that many more could be discovered in filaments and voids. 
    
    Here, we report the serendipitous discovery of a candidate of isolated and quiescent UDG, named `Zangetsu'\footnote{``ざんげつ'' in Japanese and ``斩月'' in Chinese characters. It is the name of a blade owned by the main character, Ichigo Kurosaki, of the manga ``BLEACH''. We named the discovery `Zangetsu' based on the resemblance of their shape}, in the COSMOS field (\citealp{scoville2007cosmosfield}) that could be a valuable test of the backsplash scenario. In Section \ref{sec:data}, we present the discovery of Zangetsu among the multi-band image data. We describe our galaxy modeling procedure and photometric result in Section \ref{sec:photometry}. In Section \ref{sec:results}, we measure the lower distance limits using the surface brightness fluctuations method from the image data and then investigate the potential host galaxy around Zangetsu. We discuss the possible formation mechanism for this galaxy in Section \ref{sec:formation}. 

    We adopt the AB magnitude system (\citealt{Oke1983}) with \citet{Schlafly11} Galactic extinction correction throughout this work. 
    For cosmology, we assume $H_0=70\ \rm km\ s^{-1}\ Mpc^{-1}$, ${\Omega}_{\rm m}=0.3$, and ${\Omega}_{\rm \Lambda}=0.7$.
   
\section{Discovery of Zangetsu}
\label{sec:data}

    \begin{figure*}[t!]
        \centering 
        \includegraphics[width=\textwidth]{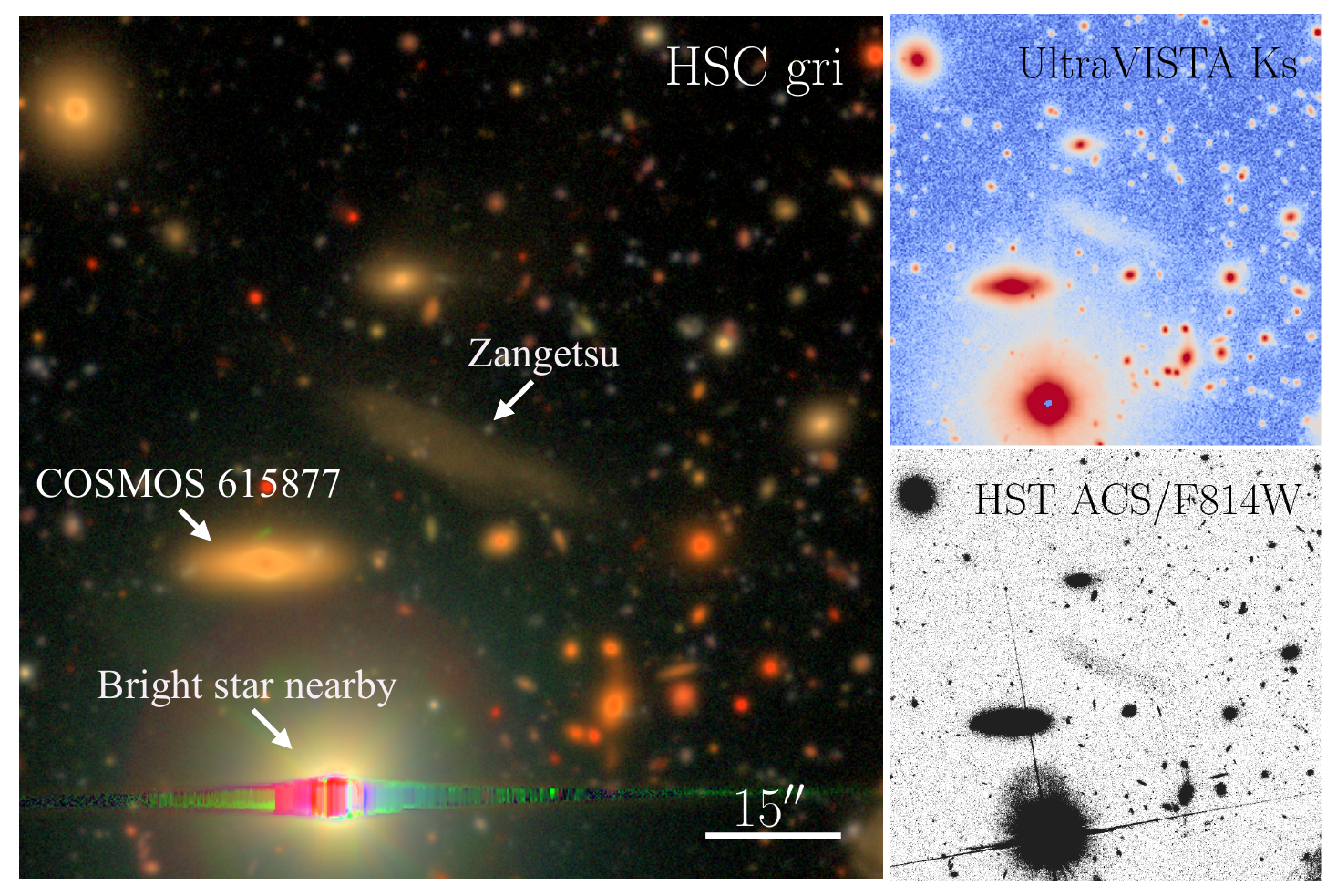}
        \caption{
        \textbf{Left}: Composite $gri$ 3-color picture of Zangetsu (\citealp{RGBimage}) from the HSC-SSP UltraDeep survey. \textbf{Upper right}: UltraVISTA near-infrared $Ks$-band image. \textbf{Bottom right}: {\it HST} ACS/F814W image.
            Zangetsu is detected in all-optical and near-infrared images of the COSMOS field. It shows a very elongated and smooth surface brightness profile, a red optical color, and a slightly distorted morphology. All cutouts here are $100^{\prime \prime}\times100^{\prime \prime}$ in size. We also label Zangetsu, the bright star, and the S0 galaxy COSMOS~615877 nearby.}
            
        \label{fig:obj_image}
    \end{figure*} 

    
\subsection{Discovery in the Hyper Suprime-Cam Survey Data}
    \label{ssec:hsc}

    We serendipitously discovered Zangetsu (also referred to as 316446 in the classic version of COSMOS 2020 catalog, \citealp{weaver2022cosmos2020}; UDG: Figure \ref{fig:obj_image}) during the visual inspection of the third public data release of the Hyper Suprime-Cam (\citealp{miyazaki2018hyper}; HSC) Subaru Strategic Program (SSP, \citealp{HSCDR3}; referred to as the HSC survey)\footnote{\url{https://hsc-release.mtk.nao.ac.jp/doc/}} in the COSMOS field while searching for other LSB targets using the interactive display of three-color images in the hscMap\footnote{\url{https://hsc-release.mtk.nao.ac.jp/hscMap-pdr3/app/}}. The HSC survey utilizes the prime focus camera of the 8.2-m Subaru telescope to conduct an ambitious $\sim 330$ nights survey of $\sim1200$ deg$^2$ of the sky in five broad bands ($grizy$, \citealt{hscfilter}), reaching a $r \sim 26.5$ mag (5-$\sigma$ point source detection limit) imaging depth in the \texttt{Wide} layer. Even more fortunately, the COSMOS field used here is part of the HSC \texttt{Deep} \& \texttt{UltraDeep} (\texttt{DUD}) layer, which reaches the impressive imaging depth of $\sim 27.4$, 27.1, 26.9 mag in the \texttt{gri} bands, making it ideal for discovering extremely LSB objects such as Zangetsu. 
    
    In the $gri$ three-color composite image (Figure \ref{fig:obj_image}), Zangetsu ($\mathrm{R.A.}=149.823^{\circ},\ \mathrm{Dec.}=1.728^{\circ}$) is next to a highly saturated bright star and an edge-on quiescent disc galaxy (S0). Just through visual inspection, it is easy to see that Zangetsu exhibits a very low central surface brightness, a very flat light distribution, an unusually elongated shape, and a red optical color. Intriguingly, while the overall surface brightness distribution of Zangetsu is smooth and featureless, it displays signs of asymmetry.

    As Zangetsu sits next to a bright star, we first investigate the possibility of any physical association with the star and confirm Zangetsu's extragalactic origin. To our best knowledge, the morphology and color of Zangetsu are distinctively different from the known cases of ejecta from stellar outbursts or nebula surrounding Young Stellar Objects (YSO; e.g., \citealp{Jochen1998yso})\footnote{For example, see the \href{https://www.ipac.caltech.edu/2mass/gallery/images_ysos.html}{2MASS Atlas Image Gallery of Young\\ Stellar Objects}}. The bright star (2MASS~J09591827+0143020 or Gaia DR3 3834532164620977152 at $\mathrm{R.A.}=149.826^{\circ},\ \mathrm{Dec.}=1.717^{\circ}$) is a $m_G=12.04$ mag, G-type main-sequence star at $\sim 255$ pc in Gaia DR3 data release (e.g., \citealt{prusti2016gaia}, \citealt{vallenari2023gaia}). We also investigate the light curve and its power spectrum from the Transiting Exoplanet Survey Satellite (TESS, \citealp{ricker2015tess}) data, finding no flare signal. Combining the large angular separation ($\sim40\arcsec$ so $\sim10000$ AU) from the star to our object, we rule out the possibility that Zangetsu has a stellar origin. We also consider the possibility that Zangetsu is a Galactic cirrus, a feature commonly seen in deep HSC images. We rule out this scenario because 1. Zangetsu's morphology does not resemble the Galactic cirrus studied in previous work (e.g., \citealp{Roman2020cirri}, \citealp{Liu2025cirrus}); 2. COSMOS field is in a low-extinction region, and no other similar structure is nearby; Although Zangetsu’s optical colors (e.g., $g-r$ vs. $r-i$) marginally overlap with the cirrus color ranges reported in recent observations (e.g., \citealp{Roman2020cirri}, \citealp{Smirnov2023GalacticCirrus}, \citealp{Liu2025cirrus}), its much smaller angular size, lack of filamentary morphology, and the absence of similar features in the surrounding region strongly argue against a Galactic cirrus source. Therefore, we conclude that Zangetsu is unlikely to be a Galactic cirrus feature.

    Based on the above analysis, we argued that Zangetsu is an extragalactic object. Before discussing Zangetsu's physical nature, we summarize its multi-wavelength detection in the next section.
\subsection{Multi-Wavelength Detection of Zangetsu}
    \label{ssec:cosmos}

    As Zangetsu is located in the COSMOS field \citep{scoville2007cosmosfield}, it has access to the full range of multi-wavelength data from a series of ground- and space-based telescopes. In particular, the sky region of Zangetsu have been also covered by the optical-NIR data from the Hubble Space Telescope ({\it HST}, \citealp{scoville2007cosmosfield}) ACS/F814W, Subaru Suprime-Cam ($BVgriz$, 12 intermediate bands, and two narrow bands), the Sloan Digital Sky Survey ({\it SDSS}, $ugriz$), CFHT MegaPrime \& WIRCAM ($uiHK_{\rm s}$), UKIRT WFCam($J$), and the UltraVISTA survey ($YJHK_{\rm s}$ and one narrow band). These data provide almost continuous wavelength coverage between 0.3 and 2.2 $\mu$m. Zangetsu is detected on \textit{all} these images with consistent position and morphology. In addition to the optical and near-infrared data, COSMOS is covered by extensive radio observations from the Very Large Array (VLA, \citealp{schinnerer2010VLAdp}), near-infrared and mid-infrared data from {\it Spitzer} (\citealp{sanders2007spitzer}), UV data from \textit{GALEX} (\citealp{guillaume2006galex}), and X-ray data from XMM-Newton (\citealp{hasinger2007xmm}). However, through visual inspection and cross-matching with the available catalogs, we find that Zangetsu is only tentatively visible in the {\it Spitzer}/IRAC1 image and undetected in all others. We provide the mosaic of Zangetsu's post-stamps from all available imaging data in the Appendix \ref{app:multiband}. 

    Zangetsu has been identified in the COSMOS2020 catalog\footnote{\url{https://cosmos2020.calet.org/catalogues/}} (\citealp{weaver2022cosmos2020}). Its ID in the classic version using \texttt{SExtractor} (\citealp{SExtractor}) is $316446$, but it is not detected in the Farmer version due to the mask for the nearby bright star. Based on the SED fitting results in the classic version of the COSMOS2020 catalog, the photometric redshift is estimated to be $z=0.17^{+0.03}_{-0.05}$ using LePHARE (\citealp{arnouts2011lephare}) and $z=0.067^{+0.015}_{-0.006}$ using EAZY (\citealp{brammer2008eazy}). The former is unreasonable as the galaxy can reach a size of $R_e\sim32\ \rm kpc$ at such a redshift, while the latter fit is also not reliable as indicated by a high $\chi^2$ value of $166.8$. The photometric redshift estimation shows that Zangetsu's LSB and quiescent nature make it difficult to measure its SED and derive accurate photometric redshifts.

    We search for the archival imaging data from other telescopes from the Mikulski Archive for Space Telescopes (MAST)\footnote{\url{http://archive.stsci.edu}} and COSMOS field Archive data cutouts\footnote{\url{https://irsa.ipac.caltech.edu/data/COSMOS/index\_cutouts.html}}. And as for the imaging data from {\it HST}, Zangetsu is detected in the Wide Field Camera 3 (WFC3) F160W data (\citealp{mowla2019cosmos}) and Advanced Camera for Surveys (ACS) F814W data (\citealp{scoville2007cosmosfield}). The top-right panel of the Figure \ref{fig:obj_image} shows the {\it HST}/F814W image. Since Zangetsu is not resolved into individual stars in this high-resolution image, it indicates that Zangetsu is unlikely to be a Local Volume dwarf galaxy.

\section{Photometric Analysis}
    \label{sec:photometry}
    
\subsection{Data Analysis and Photometric Modeling}
    \label{ssec:DataReduction}

    In this section, we describe the data analysis and photometric modeling procedures for Zangetsu. We rely on the public {\texttt{deepCoadd}} imaging data reduced by the {\texttt{hscPipe}} (e.g., \citealt{hscpipeline}). We use the deepest $gri$-band coadd images from the \texttt{DUD} layer, which has a worse PSF than those from the \texttt{WIDE} layer. However, the much deeper imaging depth benefits the study of LSB targets such as Zangetsu. The coadd images are global background-subtracted using 2d polynomials in {\texttt{hscPipe} (detailed description in Sec 3.5 of \citealt{HSCDR3})}, with precise astrometric and photometric calibration, and are analyzed by the multi-band photometric measurements procedures in {\texttt{hscPipe}}. However, due to Zangetsu's proximity to a saturated star, the highly crowded nature of deep images, and the challenges of modeling an LSB target, we implement customized procedures to ensure the reliability of the photometric analysis of Zangetsu. 

\subsubsection{Removing the Scatter Light of Nearby Star}

    Firstly, we address the issue of the nearby bright star. In {\texttt{hscPipe}}, the default empirical sky background modeling procedure uses ``superpixels'' not large enough to account for the extended light distributions around bright galaxies and stars, which leads to local over-subtraction of the sky, leaving a ``dark donut'' feature around them (e.g., see Figure 5 of \citealt{HSCDR2}). Starting with HSC PDR2, a global sky background subtraction algorithm was developed to address this issue and preserve the intrinsic light distributions of bright objects (see a detailed description in Section 4.2.6 of \citealp{HSCDR3}). We use the \texttt{deepCoadd} images with global sky subtraction that is most appropriate for studying LSB objects. However, it also means that the scattered light of the saturated star artificially boosts the ``background'' on one side of Zangetsu and contaminates its flux distribution. Given the highly extended nature of this scattered-light component (e.g., \citealt{Garate-Nunez2024}), we decided to remove it through modeling, as masking is no longer the best option.

    We generate $300\arcsec\times300\arcsec$ cutout images around the star to capture a significant fraction of the scattered light component. According to \citet{Garate-Nunez2024}, at the radius of $150\arcsec$ (or $\sim 1000$ pixels), the flux density of the PSF drops to $<10^{-6}$ of the central value. While \citet{Garate-Nunez2024} provides 2-D extended PSF models, they are designed for the \texttt{WIDE} layer images, so we use a flexible 1-D isophotal fitting procedure to create a 2-D scatter-light model to subtract. We first aggressively mask all detected objects except for the bright star on the cutout. Saturated and problematic pixels are excluded based on the bitmask plane provided by {\texttt{hscPipe}}. We then fit the bright star's 1-D surface brightness profile using the \texttt{isophote} function in \texttt{Photutils} (\citealp{photutils})\footnote{\url{https://photutils.readthedocs.io/en/stable/index.html}} from 10 to 400 pixels along the major axis with a 0.25 dex step size. While we fix the center of the stars, we allow the isophotal shapes to vary. After the fitting, we use the \texttt{build\_ellipse\_model} function to render the 2-D model and subtract it from the image. We perform this step independently in all $gri$-band images. We show the 2-D model of the star, along with the residual images in Appendix \ref{app:sky}. As the \texttt{DUD} coadd stacks many exposures with different levels of scattered light or optical ghosts around the bright star, the residual reveals complex, asymmetric features around the star's center. While this is difficult to eliminate, we mostly care about the ``gradient'' of the scattered light around Zangetsu, which is robustly accounted for through our 1-D modeling. In Appendix \ref{app:sky}, we demonstrate this using 1-D profiles of the star extracted using different configurations. We will prove later that our results are robust against the choices made here.

\subsubsection{Additional Sky Background Adjustment}
    \label{ssec:sky_subtraction}

    \begin{figure*}[t!]
        \centering 
        \includegraphics[width=\textwidth]{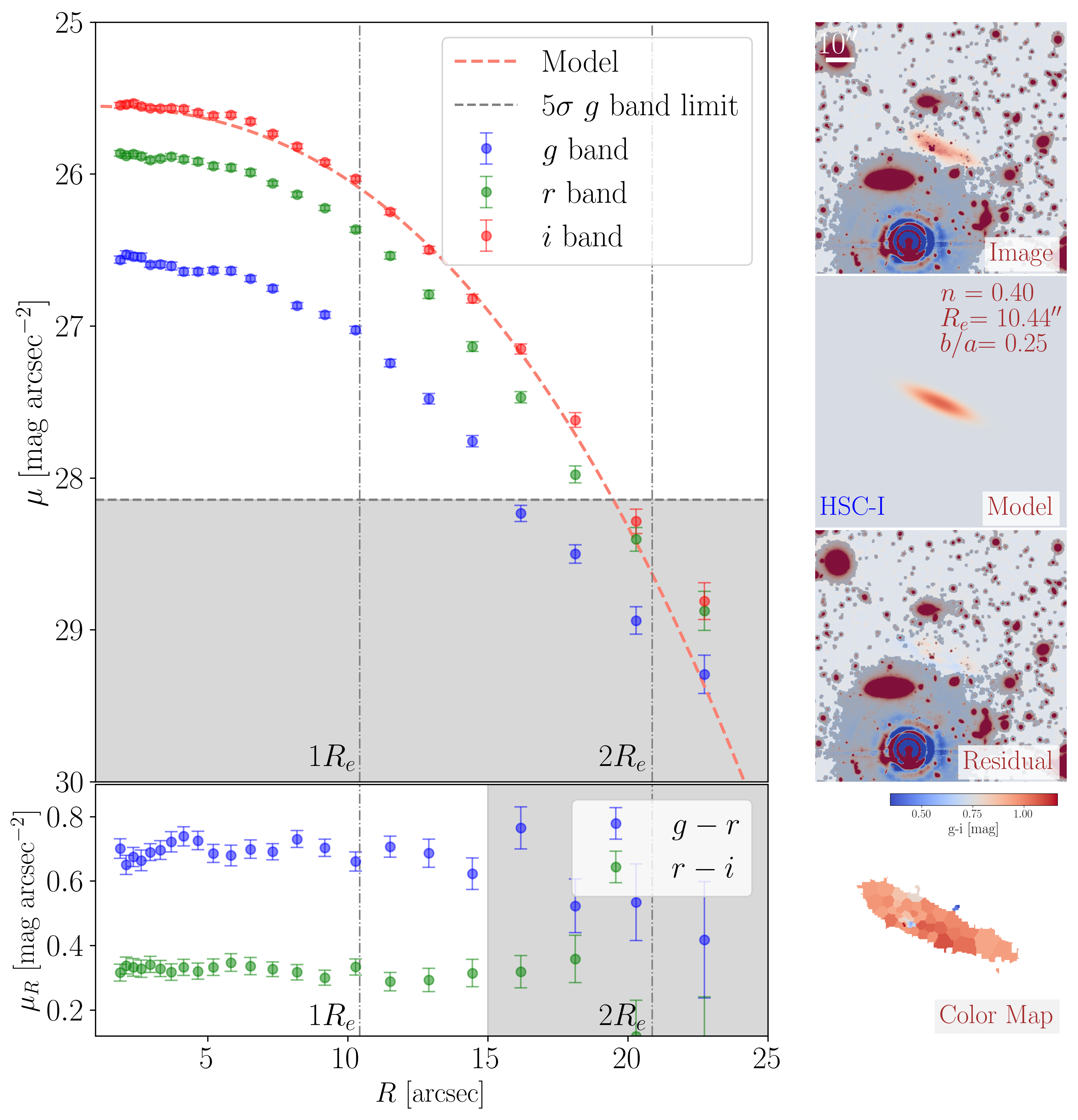}
        \caption{
            \textbf{Left panels}: Surface brightness profile and color profile of Zangetsu. The red dashed line in the top-left panel is the $i$-band \ser{} fitting result. Multiples of the effective radius are marked by vertical dashed lines. The grey shaded region indicates the $5\sigma$ surface brightness limit from $g$ band where the data points are not reliable because of low S/N. \textbf{Right panels}: The $i$-band image of the galaxy after subtracting the scattered light from the bright star (\S\ref{ssec:sky_subtraction}), \ser{} model, and residual from the \ser{} fitting. The best-fit \ser{} parameters are shown in the middle panel. All cutouts are $106\arcsec\times 106\arcsec$ in size. The spatially resolved $g-i$ color map for Zangetsu is shown in the bottom panel. Both the 1D color profile and the color map indicate that Zangetsu exhibits no significant color gradient.
            }
        \label{fig:profile}
    \end{figure*}

    As background control is critical in modeling LSB targets, we aim to ensure the sky background around Zangetsu suffers no under- or over-subtraction after removing the bright star's scattered light. In addition to the star, there is also a $z\sim 0.12$ early-type disc galaxy near Zangetsu, whose intrinsic stellar halo could affect Zangetsu's light distribution. We do not differentiate this galaxy's potential stellar halo component from the actual sky background; instead, we perform an additional local background estimation $+$ correction. We generate $106\arcsec\times106\arcsec$ cutouts around Zangetsu on the star-subtracted coadded images, aggressively mask out every detected object, and then estimate the background using the \texttt{MedianBackground} 2-D algorithm from \texttt{Photutils} with a $15\times 15$ pixel box for each band. The median background pixel values for the three bands ($gri$) are $0.0006$, $0.0003$, and $0.0003$, while the median value of the RMS map is $0.0061$, $0.0086$, and $0.0095$. This result suggests no global background-subtraction bias, and the scatter in local background could arise from unmasked extended light distributions from the star and bright galaxies. We subtract the 2-D sky models from the cutouts and re-estimate the background RMS values. The updated values $0.0020$, $0.0028$, and $0.0035$ for $gri$ bands suggest a 5-$\sigma$ surface brightness limit of $28.14$, $27.76$, and $27.51$ mag arcsec$^{-2}$ over $\sim 2.52\arcsec$ angular scale, appropriate for the reliable modeling of an LSB galaxy such as Zangetsu. Appendix \ref{app:sky} shows the 2-D sky models and the pixel distributions before and after the correction. We use these local background-corrected cutouts for modeling Zangetsu. While the choice of box size for sky background estimation has little effect on the main results, the scattered-light subtraction and background-correction steps are crucial for reliably characterizing Zangetsu's surface brightness profile. To further test the robustness of our background processing, we inject mock versions of Zangetsu into different sky backgrounds, and the sky background variation does not largely affect the result with our pipeline, as discussed in Appendix \ref{app:injection}.

\subsection{Galaxy 2-D Modeling}
    \label{ssec:galaxy_model}

    \begin{table*}[th!]
    \centering
    \setlength{\tabcolsep}{4pt}
    \begin{tabular*}{0.95\linewidth}{*{6}{c}}
    
    \toprule
     & & \multicolumn{2}{c}{\textbf{\ \ \ \ \ \ \ Basic and Morphology Parameter}}& & \\
    \hline
    RA (J2000)          & DEC (J2000)       & $n$   & $b/a$  & PA      & $R_{\rm e}$                   \\
    $149.8226^{\circ}$ & $1.7285^{\circ}$ & $0.40\pm0.01$& $0.25\pm0.01$ & $1.15\pm0.01$ & $10.44\pm0.06{\arcsec}$ \\
    9:59:17.4295 & 01:43:42.5244 & & & & \\
    \hline
     & & \multicolumn{2}{c}{\textbf{\ \ \ Photometric Parameter}}&  & \\
    \hline
    $\mu_e(g)$                & $27.09\pm0.01$ $\mathrm{mag}\ \mathrm{arcsec}^{-2}$& 
    $\mu_e(r)$                & $26.40\pm0.01$ $\mathrm{mag}\ \mathrm{arcsec}^{-2}$&
    $\mu_e(i)$                & $26.09\pm0.01$ $\mathrm{mag}\ \mathrm{arcsec}^{-2}$\\
    
    $\langle \mu_e(g)\rangle$ & $26.77\pm0.01$ $\mathrm{mag}\ \mathrm{arcsec}^{-2}$& 
    $\langle \mu_e(r)\rangle$ & $26.09\pm0.01$ $\mathrm{mag}\ \mathrm{arcsec}^{-2}$&
    $\langle \mu_e(i)\rangle$ & $25.78\pm0.01$ $\mathrm{mag}\ \mathrm{arcsec}^{-2}$\\
    
    $\mu_0(g)$                & $26.55\pm0.01$ $\mathrm{mag}\ \mathrm{arcsec}^{-2}$&
    $\mu_0(r)$                & $25.86\pm0.01$ $\mathrm{mag}\ \mathrm{arcsec}^{-2}$&
    $\mu_0(i)$                & $25.55\pm0.01$ $\mathrm{mag}\ \mathrm{arcsec}^{-2}$\\
    
    $m_g$                     & $21.13\pm0.02$ $\mathrm{mag}$  &
    $m_r$                     & $20.46\pm0.02$ $\mathrm{mag}$  & 
    $m_i$                     & $20.17\pm0.02$ $\mathrm{mag}$ \\
    
    $g-i$                     & $0.96\pm0.03$ $\mathrm{mag}$  &
    $g-r$                     & $0.67\pm0.03$ $\mathrm{mag}$  & & \\
    
    \hline
    \end{tabular*}
    \caption{
        Photometry and structural properties of Zangetsu. The \ser{} index $\mathrm{n}$, axis ratio $\mathrm{b/a}$, position angle $\mathrm{PA}$, and effective radius $\mathrm{R_e}$ are extracted from the $gri$ three-band joint \ser{} fitting. We also include the central surface brightness $\mu_0$, the average surface brightness within the effective radius $\langle \mu_e\rangle$, and the surface brightness at the effective radius $\mu_e$ for each band.}
        \label{tab:param}
    \end{table*}

        With the scattered light removed and background corrected in the $gri$ bands, we now characterize the structural and photometric properties of Zangetsu through 1-D surface brightness profile extraction and 2-D modeling using \texttt{isophote} and \texttt{AstroPhot} (\citealp{astrophot})\footnote{ \url{https://astrophot.readthedocs.io/en/latest/}}. For the 2-D fitting, we adopt the PSF model image and the variance map respectively in the $gri$ bands from the \texttt{hscPipe}. Given Zangetsu's simple morphology, we begin with a single \ser{} model and a flat-sky component. We also account for the nearby objects blended with Zangetsu by iteratively building an object mask. 

        We briefly introduce the fitting procedures here:
        
    \begin{itemize}

        \item As the first step, we fit the $gri$-band images independently with the \ser{}$+$Sky model, allowing all parameters to vary freely, including the centroid of the galaxy. The flexible 2-D models aim to achieve a smooth residual model and facilitate improving the object mask. On the residual map, we detect all objects using \texttt{Photutils.segmentation} with a lower detection threshold (3 times the median value of the sky RMS map from the residual) and a minimum number of connected pixels of $8$. We display the final object mask in Figure \ref{fig:profile}. 

        \item With a more complete object mask from the first round of fitting, we further quantify the uncertainties in our final results by performing a second round of fitting using the Markov Chain Monte Carlo (MCMC) algorithm, specifically the No-U-Turn Sampler (NUTS). Given the LSB and diffuse nature of Zangetsu, to improve the signal-to-noise ratio for the band where Zangetsu is faint, we constraint the morphological parameter including the center, axis ratio, position angle, \ser{} index, and effective radius to perform a joint fitting, but allow parameters of sky model in each band vary, since the sky background should be different between bands. We set Gaussian priors on each parameter, centered based on the initial fitting results. The priors are set to be wide enough to ensure the MCMC can effectively explore the parameter space. The MCMC is run for 5000 steps, with the first 200 discarded as burn-in and the remaining used to sample the posterior distributions. We take the mean value for each distribution as our best-fit 2D \ser{} model, and adopt the 3$\sigma$ range of the posterior distribution as the corresponding uncertainties. The corner plot of the posterior distributions is shown in the Appendix \ref{app:corner}. Note that due to the low surface brightness nature of the object, the systematic uncertainties are the same order as, or even larger than, the statistical uncertainties reported here. We evaluate the impact of each processing step in Appendix \ref{app:sky}, and further perform an injection-recovery mock test to estimate the systematic error, as discussed in Appendix \ref{app:injection}.

        \item Finally, to create a less model-dependent representation of Zangetsu and compare it with the best-fit \ser{}$+$ Sky model, we adopt the central position, axis ratio, and position angle from the best-fit 2-D model as a fixed shape to extract the 1-D surface brightness profiles in all three bands. In Figure \ref{fig:profile}, we plot the 1-D surface brightness profiles. We also display the $g-r$ and $r-i$ color profiles based on the 1-D profiles.
 
        \item Additionally, we extract the spatially resolved $g-i$ color map to further investigate the quiescent nature of Zangetsu and to rule out the possibility of an edge-on system. To prevent PSF-induced color gradients, we match the PSF of the $g$ band image to that of the $i$ band image by convolving the $g$-band image with a kernel constructed using the \texttt{psf.matching} module in \texttt{Photutils}, where a Hanning window function is applied to filter out high-frequency components during kernel generation. We apply Voronoi binning to create adaptive spatial bins with ${\rm S/N} > 20$ for the color maps, as shown in Figure \ref{fig:profile}.

    \end{itemize}

    In Figure \ref{fig:profile}, we summarize the results of the photometric modeling by showing the star-subtracted image, the best-fit $i$-band \ser{} model, the residual image,  the $gri$-band surface brightness profiles along with the best-fit \ser{} profiles, and the two color profiles. The \ser{} model accurately captures the overall surface brightness distribution of Zangetsu. However, this simple model cannot account for the sign of morphological distortion, as shown in the residual image. We will describe the photometric and structural properties of Zangetsu in the next section. Appendix \ref{app:sky} summarizes the results of alternative 2-D models with different scattered light subtraction strategies, background corrections, object masks, and model flexibility. These models demonstrate the robustness of our results.
    

\subsection{The Structural and Photometric Properties}
    \label{ssec:structure}
    
    \begin{figure*}[t!]
        \centering 
        \includegraphics[width=1\textwidth]{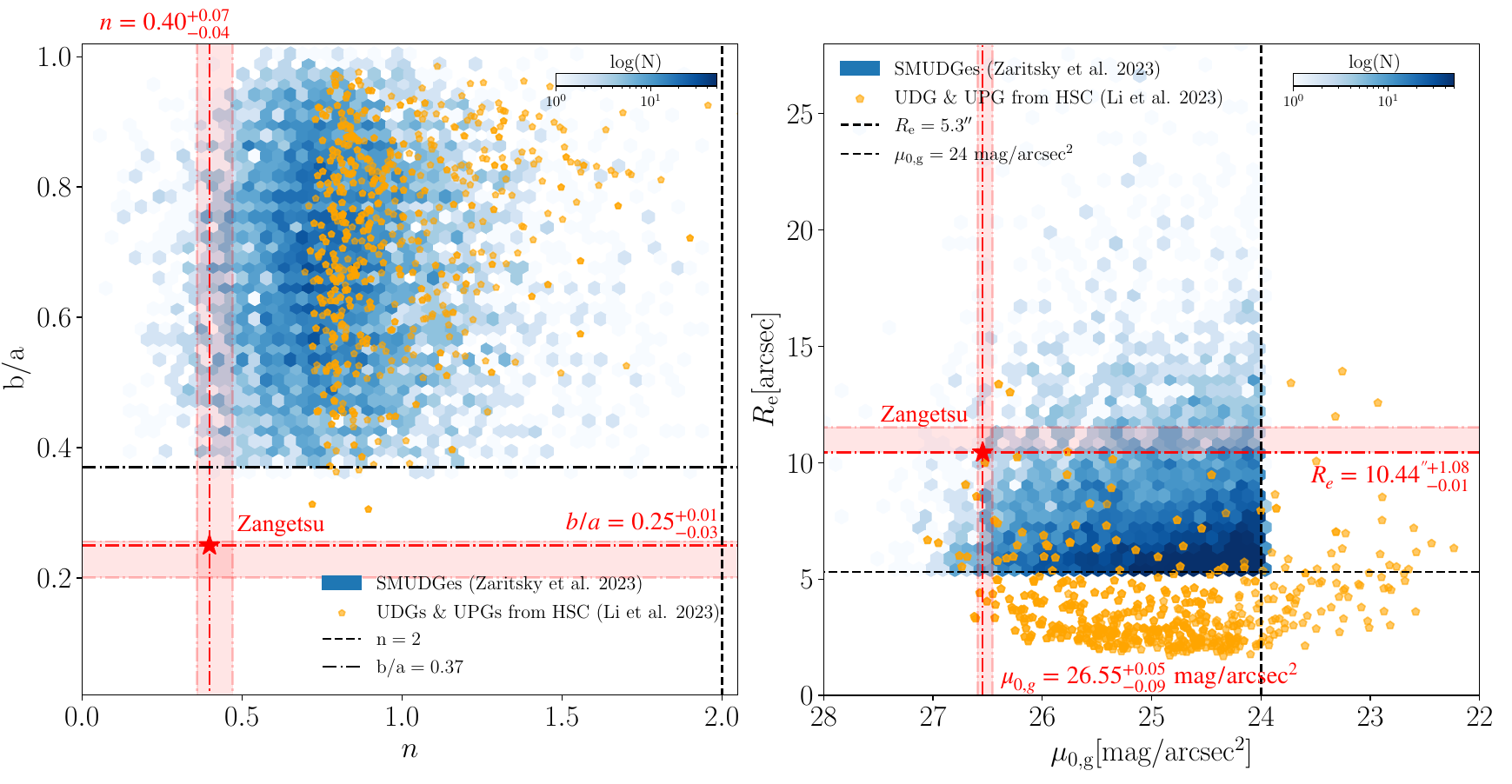}
        \caption{
            \textbf{Left panel: }
            To place Zangetsu in a broader context, we compare Zangetsu (red star) with UDGs from the SMUDGes survey (\citealt{Zaritsky2023SMUDGesV}, blue hexagonal bins) and UDGs and UPGs from Li+23 (\citealt{li2023UPG}, orange dots) on the \ser{} index $n$ vs $b/a$ 2-D plane. 
            The black dashed lines are the selection cut from the SMUDGes catalog, respectively $n<2$ and $b/a\leq 0.37$. The red dashed line marks the corresponding positions for the parameters from Zangetsu. The red shaded area represents the error range derived from the injection–recovery mock test. \textbf{Right panel: }Central surface brightness in $g$ band vs apparent effective radius in arcsec for the same samples. The black dashed lines here show the selection cut from SMUDGes catalog, which are respectively $\mu_{0,g}>24\ \mathrm{mag\ arcsec^{-2}}$ and $R_{e}>5.3\arcsec.$}
        \label{fig:morp_comp}
    \end{figure*}
    
    In Table \ref{tab:param}, we summarize the main properties of Zangetsu from the photometric modeling described above. As shown in Figure~\ref{fig:profile}, a single \ser{} profile provides a good fit to Zangetsu's $gri$-band surface brightness profiles out to nearly $2\,R_{\rm e}$, or down to a surface brightness level of $\mu_g \sim 28\ \mathrm{mag\ arcsec^{-2}}$. This limit corresponds to the $5\sigma$ surface brightness threshold in the $g$ band, estimated from the sky background, where reliable isophote extraction is still possible. According to our model, Zangetsu has a low central surface brightness ($\mu_0(g)=26.6$ mag arcsec$^2$) and a large angular effective radius ($R_{\rm e}=10.4$ arcsec). However, what makes Zangetsu unique is the highly flat surface brightness profile within $1\times R_{\rm e}$ and the very elongated shape, which are highlighted by the low \ser{} index $n\approx 0.4$ and the extremely low axis ratio ($b/a=0.25$). 
    
    To help put Zangetsu's properties in the perspective of the statistical distributions of \ser{} parameters of the known LSBs or UDGs, in Figure \ref{fig:morp_comp}, we compare Zangetsu's \ser{} index $n$, axis ratio $b/a$, effective radius $R_{\rm e}$, and $g$-band central surface brightness to the SMUDGes sample (\citealp{Zaritsky2023SMUDGesV}), the largest sample of LSB galaxies so far, as well as the UDG and ultra puffy galaxy (UPG) samples also extracted from the HSC data (\citealp{li2023UPG}), which more faithfully represent the LSB galaxy population above the luminosity-size relation. On the $\mu_0(g)$ vs. $R_{\rm e}$ plane, Zangetsu has a larger size and has a lower central surface brightness than 32\% and 6\% of the average of SMUDGes galaxies. At the same time, Zangetsu’s \ser{} index lies $\sim1.5\sigma$ below the SMUDGes median, ranking lower than 48\% of the sample and indicating a moderate but statistically meaningful deviation. However, regarding the $b/a$ distribution, Zangetsu is located outside the $b/a \geq 0.37$ selection criteria of SMUDGes, making it a unique case among the known LSB/UDG population. Putting all this together, we have reasons to believe Zangetsu is special. 
    
    On the extremely low axis ratio of Zangetsu, it is worth noting that highly elongated LSB galaxies are rare even for cases not as extreme as Zangetsu. In \citet{li2023UPG}, the authors adopted a slightly relaxed ellipticity cut ($b/a>0.35$) but only added four galaxies in the $b/a<0.37$ region. Assuming the $b/a$ distribution of the SMUDGes reflects the intrinsic one of the underlying population, we would expect $2$\% galaxies at $b/a<0.25$ by extrapolating the histogram to the $b/a<0.37$ region using a beta distribution with shape parameters $\alpha=3.83$ and $\beta=2.10$. Furthermore, the commonly adopted axis ratio cut in LSB/UDG searches (e.g., \citealt{Zaritsky2023SMUDGesV, li2023UPG, Greco2018LSB}) is justified, as the $b/a < 0.4$ regime is more heavily contaminated with edge-on disc galaxies or tidal features related to galaxy merger or interaction. In the case of Zangetsu, though, the combination of the low \ser{} index, high ellipticity, lack of infrared detection (Figure \ref{fig:multiband_image}) and absence of color gradient on the spatially resilved $g-i$ color map makes it unlikely to explain Zangetsu as an ``edge-on disk'' scenario since the profile deviates significantly from an exponential disk. And, as will be discussed later, Zangetsu itself is unlikely to be a tidal feature as in the ``Sumo Puff'' case (\citealt{greco2018sumo}): the redshift of the nearby galaxy, the lack of other signatures of tidal interaction, and the relative configuration between the galaxy and Zangetsu all make it challenging for such a picture. 

    At the same time, Zangetsu does show a distorted morphology, as evidenced by the asymmetric light distributions in all optical bands and the residual image of the 2-D \ser{} model (see the bottom-right panel of Figure \ref{fig:profile}). Also, despite the low S/N, there is evidence that Zangetsu's outer surface brightness profile falls more sharply than the \ser{} model, suggesting an even lower \ser{} index for the outer part, which could be related to the physical process that also causes the morphological distortion. 
    
    Both the overall optical colors based on the 2-D models and the 1-D color profiles suggest that Zangetsu has a very red optical color ($g-i\approx 0.97\ \rm mag$). The 1-D color profiles reveal no clear sign of a $r-i$ and $g-r$ color gradient. Consistently, the spatially resolved $g-i$ color map also reveals no significant color gradient, reinforcing the picture that Zangetsu is dominated by an old stellar population with no recent star formation. Additional evidence for its quiescent nature comes from its non-detection in the GALEX UV image. W Using the GALEX NUV detection limit ($M_{\rm AB}=26\ \rm mag$), we estimate an upper limit on the star formation rate before dust correction following \citet{Kennicutt1998SFR}. The derived upper limits are $\sim7.9\times10^{-6}\ M_{\odot},\mathrm{yr^{-1}}$ at $z=0.006$ and $\sim2.5\times10^{-3}\ M_{\odot},\mathrm{yr^{-1}}$ at $z=0.1$, indicating that Zangetsu remains consistent with a quiescent system regardless of the assumed distance.
    
    Although the exact redshift of Zangetsu is unknown, assuming it at $z\sim 0.1$, we perform spectral energy distribution (SED) modeling based on the 33-band aperture photometry or the upper limits in the COSMOS2020 catalog (\citealp{weaver2022cosmos2020}) using the {\tt CIGALE} algorithm (\citealt{Boquien2019cigale}) supported by the large language model, or LLM-based agent {\tt Mephisto} (\citealt{Sun2024mephisto}). We will describe the details in Appendix \ref{app:sed}. An old stellar population with no ongoing star formation can explain Zangetsu's optical-NIR SED with a stellar mass of $M_\star \approx 1.35\times10^9\, M_{\odot}$ after an aperture correction based on the $i$-band. Assuming Zangetsu is at $z=0.006$ will reduce the stellar mass to $4.37\times10^6\, M_{\odot}$, but will not qualitatively change its quiescent nature. 
\section{Distance \& Environment}
    \label{sec:results}

\begin{figure*}
    \centering
    \includegraphics[width=0.8\linewidth]{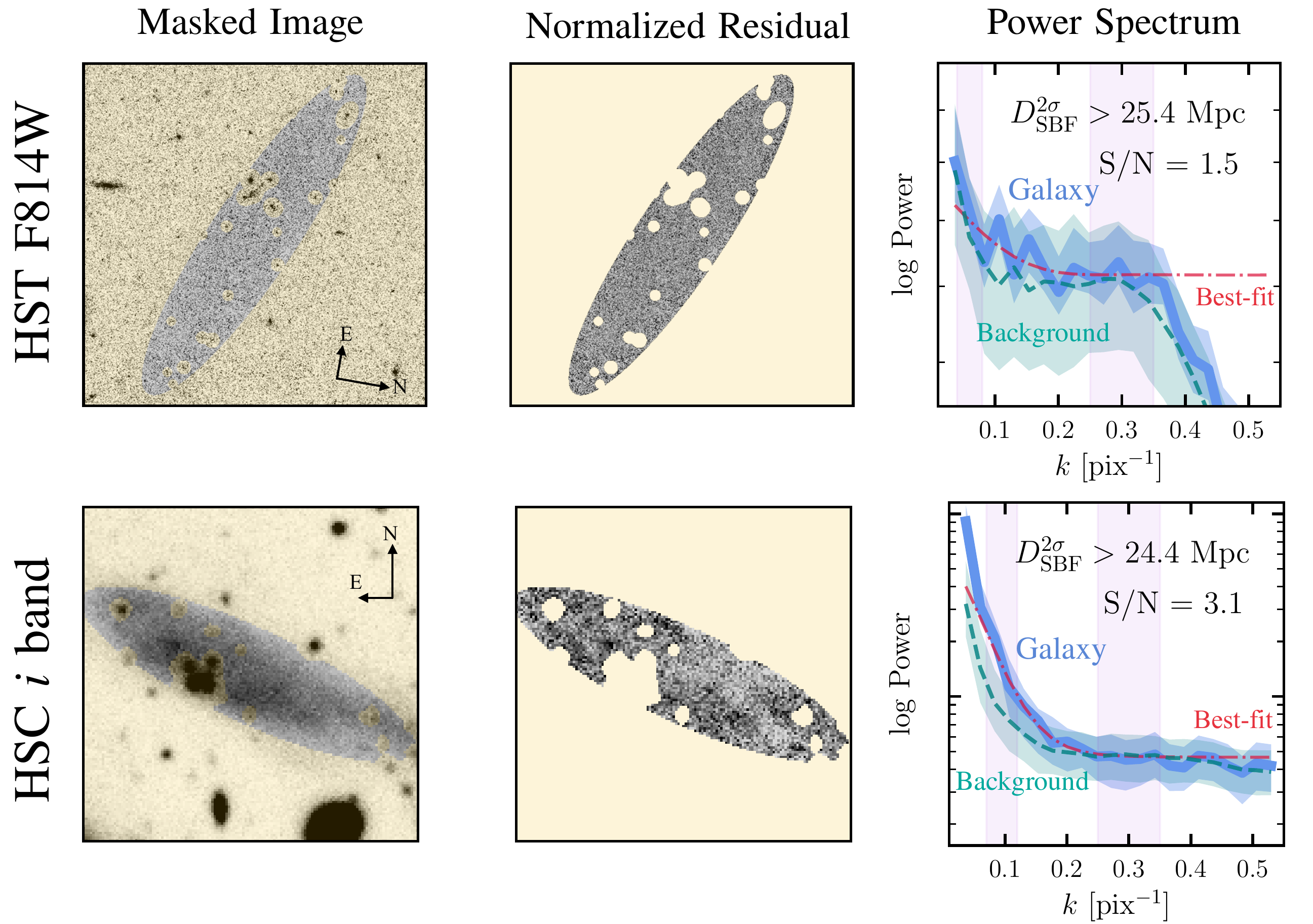}
    \vspace{-1em}
    \caption{
        Distance lower limits from surface brightness fluctuation measurements using HST F814W (top panels) and HSC $i$-band (bottom panels) images. The left panels show the images and the masks overlaid in yellow. We only include relatively high surface brightness regions ($<1.2\ r_{\rm eff}$) in our SBF measurement. The background galaxies are masked. We fit a smooth model to the galaxy, subtract it from the image, and divide its square root to normalize the residual (middle panels). Then, we calculate the power spectrum of the residual and fit it with a combination of the PSF power spectrum and a constant white noise. The same is done for blank background fields, shown as the teal band. For both HST and HSC data, we do not detect significant SBF signals, as indicated by low SBF S/N. Therefore, we report the distance $2\sigma$ lower limits to be $D_{\rm SBF}^{\mathrm{HST}} > 25.4\ \mathrm{Mpc}$ and $D_{\rm SBF}^{\mathrm{HSC}} > 24.4\ \mathrm{Mpc}$. 
    }
    \label{fig:sbf}
\end{figure*}

\subsection{Surface Brightness Fluctuation Distance}
    \label{ssec:sbf}
    
    We attempt to measure the distance to Zangetsu using the surface brightness fluctuation (SBF, \citealt{Tonry1988}; see \citealt{Cantiello2023} for a review) technique. SBF measures the pixel-to-pixel fluctuations due to Poisson noise in the number of stars per pixel. As the distance increases, the average number of stars per pixel increases, reducing pixel variance. This fact makes it possible to measure the distance to semi-resolved galaxies using only imaging data. Nevertheless, the SBF signal also depends on the stellar populations and is often calibrated against optical colors. For dwarf galaxies, the compact background sources and globular clusters also contribute significant fluctuation signals and must be masked and corrected. In this section, we measure the SBF distance to Zangetsu in a similar way to \citet{li2024hedgehog}, and we also refer interested readers to \citet{Carlsten2019}, \citet{Greco2021}, and \citet{Kim2021} for more details regarding SBF measurement for dwarf galaxies. 

    SBF measurements require high-resolution, deep data. Among the datasets available for Zangetsu, HSC \texttt{UltraDeep} (hereafter HSC) and HST/ACS F814W data are the best for SBF measurement. The HSC data are deeper than HST F814W but have a much lower resolution. We use HSC $i$-band data for SBF because the seeing is typically better and the intrinsic SBF signal is stronger in the $i$ band. The HSC $i$ filter overlaps significantly with the HST F814W filter, and SBF--color calibrations exist for both filters \citep[e.g.,][]{Blakeslee2010,Carlsten2019,Kim2021,Cantiello2023}.

\subsubsection{HST} 
    \label{sssec:hst_sbf}

    We first measure SBF distance using the HST F814W data. Instead of using the standard HST coadd image from MAST, we use \texttt{AstroDrizzle}\footnote{\url{https://drizzlepac.readthedocs.io/en/latest/}} \citep{Drizzle2002, Drizzle2010} to generate the drizzled image from the flat-fielded frames (\textit{flc} files) with a \texttt{Lanczos3} kernel and no rotation to preserve the white noise property at wavenumber $k \lesssim 0.35\ \rm{pixel}^{-1}$ while introducing slight blurring and other artifacts \citep[e.g.,][]{Mei2005}. Unlike \texttt{AstroDrizzle}'s default \texttt{square} kernel, this procedure introduces little correlated noise to facilitate the separation of the real SBF signal in the power spectrum. The drizzled image has a total exposure time of 2028.0 s and a pixel scale of $0\farcs05$ per pixel (see the top-left panel in Figure \ref{fig:sbf}). We approximate the PSF at the location of Zangetsu using an unsaturated star nearby.

    We fit a smooth \ser{} model to the star-removed image (top-left panel in \ref{fig:sbf}), subtract it from the image, and normalize the residual by dividing it by the square root of the \ser{} model (top-middle panel in \ref{fig:sbf}). To mask out background galaxies and globular clusters that also contribute to the measured SBF signal, we take the approach in \citet{Carlsten2019} and mask objects brighter than a certain absolute magnitude threshold $M_{\rm F814W}$. In practice, a threshold of $-4 < M_{\rm F814W} < -6$ could remove most globular clusters and bright background galaxies while retaining the asymptotic giant branch stars and the brightest red giant branch stars. Because the distance to Zangetsu is unknown, we assume Zangetsu is at 25 Mpc to make the mask, which is approximately the SBF distance limit of the HSC \texttt{UltraDeep} data \citep{Greco2021}. Under this distance, we set $M_{\rm F814W} = -4.8$ mag, which effectively masks out the background sources in the HST F814W data. To boost the signal-to-noise ratio, we also mask out the galaxy outskirts where the surface brightness drops below $f_{\rm mask}$ times the central surface brightness, where $f_{\rm mask} < 1$ controls the radial range of the mask. The mask is highlighted in yellow in the left and middle panels in \ref{fig:sbf}. 

    We calculate the power spectrum of the masked and normalized residual image and compute the azimuthal averaged power spectrum, shown as the blue solid line in the top-right panel in Figure \ref{fig:sbf}. We then fit the 1D power spectrum to the PSF power spectrum and a constant white-noise level. The coefficient of the PSF component is the measured SBF signal. Since unmasked background sources could also contribute to the measured SBF signal, we account for this by measuring SBF in background blank fields surrounding Zangetsu. For a given blank field, we normalize it using the same \ser{} model of Zangetsu, do masking using the same threshold, and calculate and fit the power spectrum. The teal-shaded region in Figure \ref{fig:sbf} shows the power spectra of the blank fields, representing the contribution from unmasked background sources. The SBF signals from blank fields are then subtracted from the measured SBF signal of the galaxy. 

    The uncertainty of the resulting SBF signal is estimated by doing Monte Carlo simulations by uniformly sampling a few key parameters \citep{Cohen2018}: the wavenumber range used for power spectrum fitting, $k_{\rm low} \sim \mathcal{U}(0.04, 0.08)\ \rm{pixel}^{-1}$, $k_{\rm high} \sim \mathcal{U}(0.25, 0.35)\ \rm{pixel}^{-1}$ (shown as purple vertical bands in Figure \ref{fig:sbf}), and the radial mask parameter $f_{\rm mask} \sim \mathcal{U}(0.4, 0.5)$ (corresponding to $1.2 - 1.3\ r_{\rm eff}$). We choose these wavenumber ranges to ensure that the fitting is not affected by correlated noise at large $k$ and by the \ser{} fitting residuals at small $k$. We perform 100 Monte Carlo runs to estimate the SBF signal-to-noise ratio (S/N) using the median SBF signal (after subtracting the contribution from the background) divided by its standard deviation. 

    It is evident from Figure \ref{fig:sbf} that the power spectrum of Zangetsu is virtually identical to that of the background fields, and the SBF S/N is very low ($\rm S/N = 1.5$), indicating that no significant SBF from Zangetsu is detected in the HST F814W image. Therefore, we can only derive a lower bound on the distance from SBF. We convert the measured SBF signals to distances using the SBF--color calibration in \citet{Blakeslee2010}. We note that the calibration in \citet{Blakeslee2010} requires $\rm F475W - F814W $ color, whereas we only have HSC $g-i$ color (Table \ref{tab:param}). To obtain the required color, we generated synthetic photometry using the simple stellar population models from MIST \citep{Choi2016}. We derived $$\mathrm{F475W - F814W} = 0.079\ (g-i)^2 + 0.995\ (g-i) + 0.017$$ with a residual less than 0.01 mag. Given the low SBF S/N, we report the $2\sigma$ lower limit of the SBF distance: 

    \begin{equation}
        D_{\rm SBF}^{\mathrm{HST}} > 25.4\ \mathrm{Mpc}.
    \end{equation}
    
    This corresponds to a radial velocity of $cz > 1428\ \mathrm{km\ s^{-1}}$ based on the Cosmicflows-3 model \citep{Kourkchi2020}. 

\subsubsection{HSC} 
    \label{sssec:hsc_sbf}

    We repeat the SBF measurement on the HSC $i$-band coadd image, where the \texttt{Lanczos5} kernel adopted by \texttt{hscPipe} ensures that the noise at small wavenumbers is uncorrelated. Thus, we use the same star-removed image as in \ref{sec:photometry} and the PSF model at the location of Zangetsu by \texttt{hscPipe} for SBF. The much deeper $i$-band HSC image reveals features that cannot be fitted well by the \ser{} model, whose large-scale residual (see the bottom-right panel in Figure \ref{fig:profile}) will leak into the power spectrum at low wavenumber and bias the SBF measurements \citep[e.g.,][]{Blakeslee1999, Cantiello2005}. Therefore, we improve the galaxy model by fitting a smooth non-parametric model to the \ser{} residual using \texttt{sep} \citep{SExtractor, PythonSEP}. 

    We take the \ser{} model derived in Section \ref{ssec:structure} and subtract it from the image to get the residual image. Then, we detect sources with a $1\sigma$ threshold and a 4-pixel mesh to mask compact sources in the residual image aggressively. We also construct a background model from this masked residual image using a 5-pixel mesh and a 2-pixel median kernel\footnote {Based on our tests, slight changes in the mesh and kernel sizes do not significantly alter the distance lower limit.}. We iterate this process twice to ensure the residuals are captured in our background model. Ultimately, we combine the \ser{} and background models to be the final galaxy models. This smooth model better describes the asymmetric morphology of Zangetsu and shows little power at the PSF scale, as indicated by the normalized residual used to measure the SBF signal (bottom-middle panel in Figure \ref{fig:sbf}).

    Because the HSC data is significantly deeper than HST, we lower the masking threshold to $M_{i} = -4.3$ mag to mask background sources better. The radial mask parameter is updated to $f_{\rm mask} \sim \mathcal{U}(0.3, 0.5)$ (corresponding to $1.2 - 1.5\ r_{\rm eff}$). We also increase the lower limit on the wavenumber to $k_{\rm low} \sim \mathcal{U}(0.07, 0.12) \ \rm{pixel}^{-1}$ to mitigate the impact of large-scale residuals from galaxy modeling. Other settings are identical to Section \ref{sssec:hst_sbf}. The resulting SBF results are shown in the bottom-right panel in Figure \ref{fig:sbf}. It has $\rm S/N = 3.1$, still below the S/N threshold ($\approx 5$, see \citealt{CarlstenELVES2022}) for a reliable SBF distance. Using the SBF -- color calibration in \citet{Carlsten2019}, we derive the $2\sigma$ lower limit of the SBF distance:

    \begin{equation}
        D_{\rm SBF}^{\mathrm{HSC}} > 24.4\ \mathrm{Mpc}.
    \end{equation}

    The lower limit of Zangetsu's distance from HSC coincidentally agrees with the HST result because SBF is sensitive to depth and resolution. HSC prevails in depth but lacks resolution. 

\subsection{Luminosity--Size Relation of Zangetsu}
    \label{ssec:Compare}

    \begin{figure*}[t!]
        \centering 
        \includegraphics[width=0.8\textwidth]{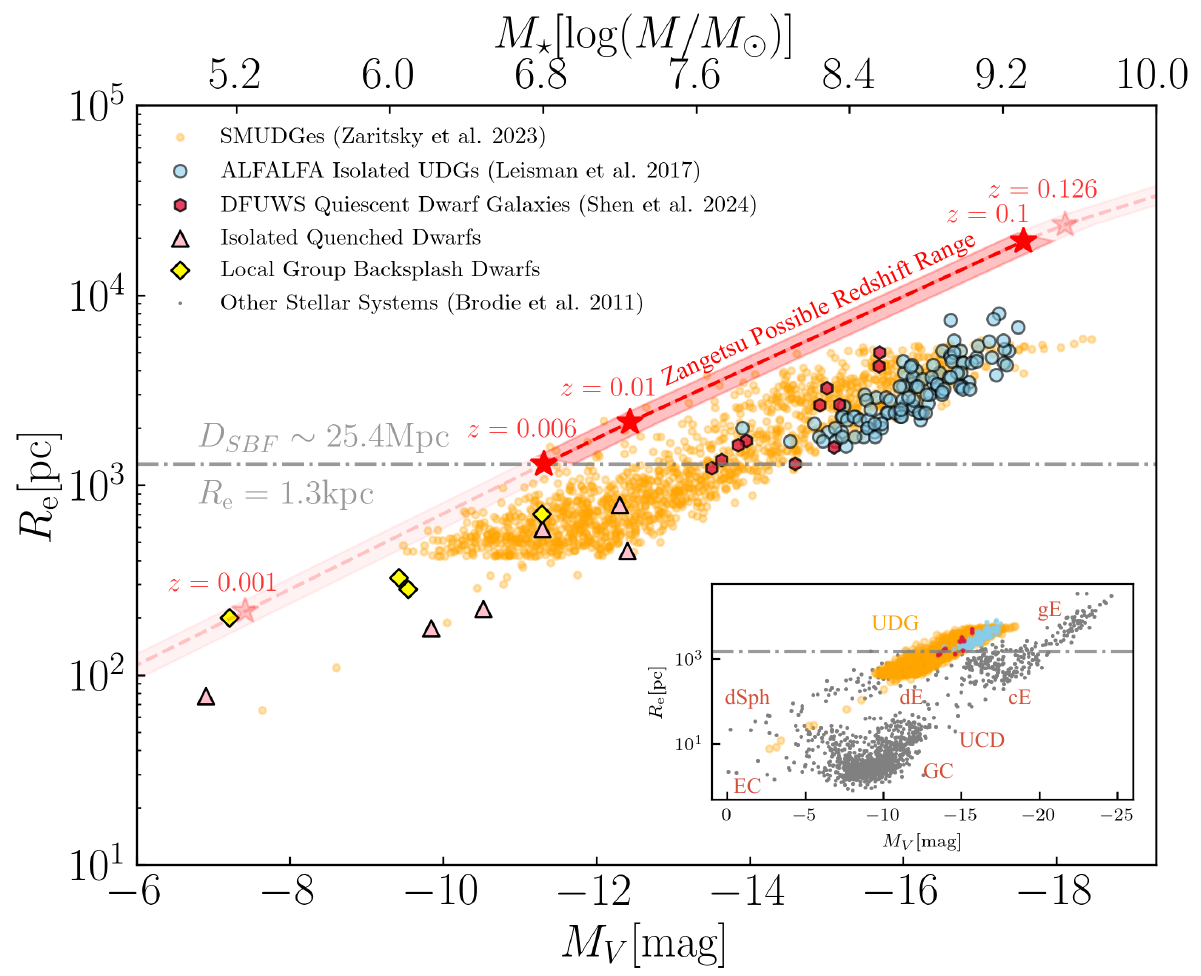}
        \caption{
            Luminosity $M_V$ and half-light radius $r_h$ parameter space for UDGs sample from various surveys (\citealp{Zaritsky2023SMUDGesV}, \citealp{leisman2017alfalfaUDG}, \citealp{shen2024first}) and Zangetsu. The horizontal gray dash-dotted line indicates the physical size of Zangetsu corresponding to the lower distance limit estimated from the SBF method. We estimate the reasonable redshift range for Zangetsu, of which the lower limit is from SBF and the upper limit from size comparison, shown as the deeper red dashed line. The red shaded region shows the error range estimated from the injection-recorvery mock test. The top axis shows the corresponding stellar mass, estimated using the $g-i$ color of Zangetsu and a color--$M_\star/L$ relation from \citet{bell2003MLratio}. Note that the data points from the ALFALFA survey are estimated using the ALFALFA flow model distances (\citealp{Haynes2011ALFA}) and surface brightness profiles. They are on average $\sim$0.5 mag brighter than aperture magnitudes, but this does not affect our conclusions. The inset plot in the bottom right shows a zoom-out view of this figure for other stellar systems (\citealp{brodie2011relationships}), including GCs (globular clusters), UCDs (ultra-compact dwarfs), ECs (extended clusters), dwarf spheroidals (dSphs), dwarf ellipticals (dEs), compact ellipticals (cEs), and giant ellipticals (gEs). Additionally, we compare Zangetsu and other isolated quenched dwarfs and local group backsplash dwarfs (\citealp{casey2023Blobby,sand2022tucanaB, Polzin2021cosmosdw1, Karachentsev2015kks3, Makarov2012kkr25, li2024hedgehog, Sand2024Sculptor}). This figure confirms that Zangetsu is an outlier in size, regardless of its redshift.}
        \label{fig:L_size}
    \end{figure*}

    The luminosity--size relation is the most commonly adopted empirical tool when discussing the structure of galaxies (e.g., \citealp{Kormendy2009structural}), including UDGs and other LSB populations. Therefore, understanding Zangetsu's location relative to other known UDGs on this crucial scaling relation may help us understand its physical nature and origin. However, the photometric redshift of Zangetsu is not reliable enough to constrain its distance, while the SBF measurement only provides a $D_{\rm SBF} > 25.4\ \mathrm{Mpc}$ lower limit. So we assume Zangetsu could span a wide range of redshifts and show it as a ``track'' on the luminosity-size relation in Figure \ref{fig:L_size}. 

    Here, we compare Zangetsu with a variety of UDGs and relevant dwarf galaxies with distance or redshift measurements, including the SMUDGes sample (\citealp{Zaritsky2023SMUDGesV}), the ALFALFA HI isolated UDGs \citep{leisman2017alfalfaUDG}, the quenched diffuse dwarf from the Dragonfly Ultrawide Survey (\citealt{shen2024first}), and the known candidates of isolated quenched dwarfs (e.g., \citealp{casey2023Blobby,sand2022tucanaB, Polzin2021cosmosdw1, Karachentsev2015kks3, Makarov2012kkr25, li2024hedgehog, Sand2024Sculptor}) \& Local Group backsplash dwarfs (e.g. \citealp{McConnachie2012localdwarf}). These samples represent the diversity of possible environments and other physical mechanisms that drive the evolution of UDGs. During the comparison, we neglect the slight difference between the HSC and SDSS filters and convert the $g$-band magnitude of Zangetsu to $V$-band and $B$-band using the empirical color transformation relation: $V=g-0.5784(g-r)-0.0038$ and $B=g+0.3130(g-r)+0.2271$\footnote{\url{https://www.sdss3.org/dr8/algorithms/sdssUBVRITransform.php\\\#Lupton2005}}. We also show the corresponding stellar mass to the $M_V$ based on the $B-V$--$M_\star/L$ relation from \citet{bell2003MLratio}.
    Taking the angular size and apparent magnitude of Zangetsu, the red dashed line shows the location of Zangetsu at different redshifts in Figure \ref{fig:L_size}, which presents a striking first impression: Zangetsu is almost certainly an outlier of the luminosity-size relation, even when compared to the most extreme UDGs known today \emph{regardless of its exact distance}. At the lower limit of the distance ($z\sim 0.006$ on the Hubble flow), Zangetsu's half-light radius ($R_{\rm e}$) is already close to the $R_{\rm e}>1.5$ kpc criteria defined by \citet{van2015forty}, placing it in the ``classical UDG'' regime. And from $z\sim 0.006$ to $z<0.1$, Zangetsu's luminosity--size ``track'' sits comfortably on top of the known UDG population, including the isolated or the gas-rich ones. Within this redshift range, Zangetsu would have $M_\star\sim10^7-10^9M_{\odot}$, consistent with being a dwarf galaxy (e.g., \citealt{Bullock2017}). 
    
    At $z=0.1$, Zangetsu would have a physical $R_{\rm e}$ of 19 kpc at the luminosity of $M_V=-17.6$ mag, making a singular case in the entire luminosity-size parameter space of all nearby galaxies \citep{brodie2011relationships}, as highlighted by the zoom-out panel in Figure \ref{fig:L_size}. Although that would make Zangetsu a spectacular discovery, we speculate that this possibility is unlikely. Note that the $z=0.126$ symbol highlights the redshift of the early-type disc galaxy near Zangetsu. Assuming Zangetsu is physically associated with it, Zangetsu would become a 47 kpc long ``tidal feature'' from end to end. Tidal tails or shells with even longer sizes are known (e.g., NGC~3785: \citealt{Watts2024}; NGC~474: \citealp{Bilek2022NGC474shell}; NGC~4651: \citealp{Foster2014NGC4651shell}), but they all display a clear physical connection with a visibly disturbed host. This is not the case for Zangetsu. Next, we will investigate Zangetsu's environment and possible host galaxy and discuss its physical origin in Sec. \ref{sec:formation}.


\subsection{Environment}
    \label{ssec:environment}

\begin{figure*}[t!]
    \centering 
    \includegraphics[width=\textwidth]{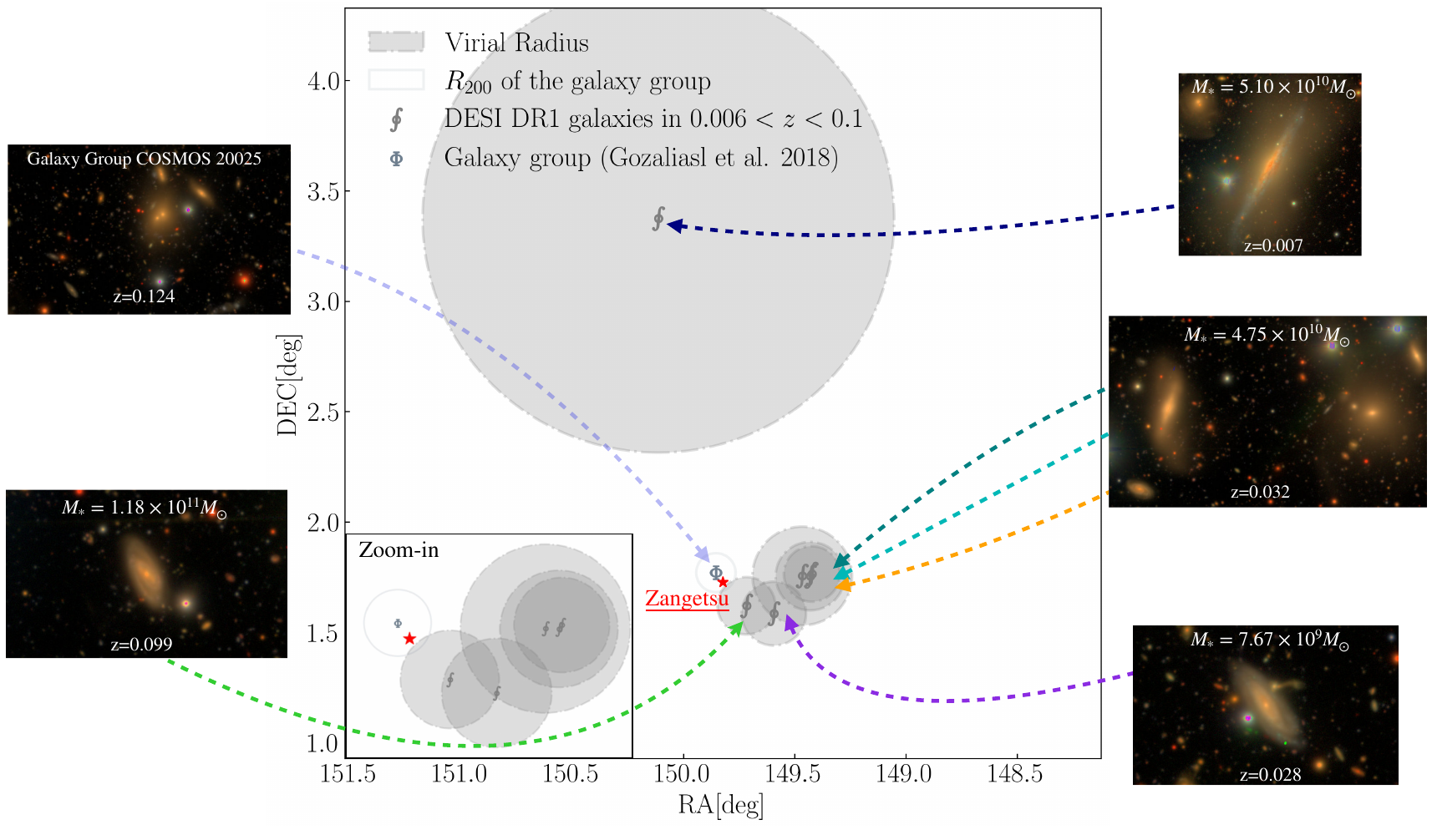}
    \caption{
        Environment of Zangetsu and potential host galaxies. We show the position of nearby galaxies or groups that meet our criteria for potential hosts (see Sec. \ref{ssec:environment} for details). Gray circles indicate the virial radius for individual galaxies and $R_{200c}$ for the galaxy group. We also show HSC $gri$-composite images for each object. The cutout images have angular sizes of $80\arcsec\times80\arcsec$ for the galaxy in $z=0.007$, $100\arcsec\times60\arcsec$ for the galaxy pair in $z=0.032$ and the galaxy group in $z=0.124$, and $50\arcsec\times30\arcsec$ for the remaining galaxies. As Zangetsu is not within the virial radius range of any potential host galaxies shown in the figure, we conclude it is very likely an isolated UDG.
        }
    \label{fig:environment}
\end{figure*}

    The environment is a critical factor in the formation and evolution of dwarf galaxies, especially low-surface-brightness galaxies. Here, we investigate the environment of Zangetsu, focusing on the search for galaxies it could have interacted with and potential dark-matter halo hosts. The successful identification of a host galaxy/halo would provide a more accurate redshift to estimate Zangetsu's physical size and stellar mass and offer clues to its possible formation channels.

    Firstly, we examine the possibility that Zangetsu is interacting with COSMOS~615877 at $z=0.126$, the nearby galaxy visible in Figure \ref{fig:obj_image} -- a massive ($M_{\star}\sim10^{11.2}\ M_{\odot}$, with a central velocity dispersion of $\sim 200$ km/s based on the Wisconsin PCA-based value added catalog\footnote{\url{https://www.sdss4.org/dr14/spectro/galaxy_wisconsin/}}), quiescent, edge-on S0 galaxy. At that redshift, Zangetsu would be well within its virial radius. In \citet{greco2018sumo}, the authors uncovered an interesting case, the ``Sumo Puff'', using HSC images. Sumo Puff sits next to SDSS J1449-0042 at z=0.00431, a post-merger dwarf galaxy with a stellar mass of $M_\star \sim 1.3\times 10^9\, M_{\odot}$. \citet{greco2018sumo} proposed that Sumo Puff could be a galaxy-size tidal feature around SDSS J1449-0042 with a linear physical dimension of $\sim 11$ kpc. The authors found tentative evidence of the LSB ``bridge'' between the two galaxies and confirmed the feasibility of their relative configuration using N-body galaxy merger simulations. The tidal feature scenario provides a convenient explanation for Sumo Puff's asymmetric light distribution, its ``clumpiness'', and its similar optical colors with SDSS J1449-0042. At a significantly higher redshift than SDSS J1449-0042, the ``tidal feature'' scenario would make Zangetsu have a linear physical size of 47 kpc, quite extreme as a ``galaxy size'' tidal feature. Large-size tidal tails or shells are not uncommon (e.g. \citealp{Watts2024, Bilek2022NGC474shell, Foster2014NGC4651shell}). Still, the known cases always show clear evidence of direct interaction when their sizes become comparable to those of the host galaxy, either through the continuity of the surface brightness distribution or through a geometric configuration that can be easily tied to gravitational interaction. Unlike SDSS J1449-0042, COSMOS~615877 does not show a post-merger or interacting signature. Even on the residual map using the star-subtracted image, we see no evidence of any surface brightness connection between Zangetsu and COSMOS~615877. And Zangetsu's almost ``linear'' morphology \& its $\sim30$ degrees angle relative to COSMOS~615877's edge-on disc distinguish it from the known tidal features. At $z=0.126$, Zangetsu would be $M_\star \sim10^{9.5}\ M_{\odot}$, $\sim 1/50$ of COSMOS~615877's stellar mass. Zangetsu also has a much bluer optical color than COSMOS~615877 ($g-r\approx 1.1$, $g-i\approx 1.6$). Without a spectroscopic redshift or a more comprehensive simulation-based study, we cannot rule out the possibility that Zangetsu is an extremely diffuse galaxy distorted during the early stage of a 1:50 merger. However, based on the above discussion, we deem the chance of this scenario to be very small. Therefore, we will explore Zangetsu's surroundings further to identify candidates of its host galaxy within $0.006<z<0.1$, using the lower limit constrained by the SBF method in Sec. \ref{ssec:sbf}.

    Next, we search for the possible host of Zangetsu in the COSMOS field using available group catalogs and galaxies with known spectroscopic redshifts, whose distance to Zangetsu in projection is smaller or close to their ``halo boundary'', such as the $R_{200c}$ -- the radius of a sphere where the mean density is $200$ times the critical density of the universe. Using the COSMOS Chandra X-ray group catalog (\citealp{gozaliasl2019chandra}) as the reference, we find only one possible group including Zangetsu within its $R_{200c}$, COSMOS~20025, at $z\sim 0.124$ with a halo mass of $M_{200c}\sim4.4\times10^{13}\ M_{\odot}$ and a $R_{200c}\sim0.7$ Mpc. Although this redshift lies slightly outside our selection criterion, we still consider the possibility that Zangetsu is physically associated with COSMOS20025, given their close projected distance. if so, Zangetsu would be a $R_{\rm e}\sim 22.5\ \rm kpc$ LSB galaxy at the projected distance of 0.6$\times R_{200c}$. It might be worth noting that the COSMOS~20025 group is at roughly the same redshift as the COSMOS~615877, which could make the massive S0 galaxy next to Zangetsu a group member. However, as discussed above, the large physical size implied and the lack of other interaction signatures in the ultra-deep HSC images make it challenging to identify COSMOS~20025 as Zangetsu's host.

    Then, we further broadened our search to all galaxies with $0.006<z<0.1$ and spectroscopic redshifts. Due to the lack of reliable $R_{200c}$ estimations for individual galaxies, we begin the search using a very generous 1 Mpc criterion to be thorough. However, this threshold value is close to the $R_{200c}$ of a massive galaxy cluster, much larger than the physical halo boundaries of most isolated galaxies or small galaxy groups. Taking advantage of the recently released Data Release 1 (DR1) of the Dark Energy Spectroscopic Instrument (DESI, \citealp{DESIDR1}), which includes more complete redshift measurements for $z<0.1$ galaxies than the SDSS catalog, we find seven possible galaxies and galaxy pairs with reliable redshift (\texttt{zwarn>0}) meet this criterion, after also set a stellar mass criterion of $\log M_*/M_{\odot}>9$ to exclude dwarf galaxies (\citealp{Bullock2017}). In Figure \ref{fig:environment}, we show the HSC images and their relative locations to Zangetsu. 

    Most of these candidates appear to be spiral or disk galaxies, so we expect them to live in dark matter halos with $R_{200c} < 1$ Mpc. Therefore, to evaluate their potential physical relations to Zangetsu, we take their stellar mass estimations from the \texttt{FastSpecFit} catalog (\citealp{FastSpecFit}, Moustakas et al.\ in prep) based on DESI DR1 and the stellar-to-halo mass relation (SHMR) from \citet{Rodriguez2017SHMR}:
    
    \begin{align}
            \langle\log\ M_{\rm vir}(M_*)\rangle = & \log\ M_1 + \beta\log\ (\frac{M_*}{M_{*,0}})\nonumber\\
            &+\frac{(\frac{M_*}{M_{*,0}})^{\delta}}{1 + (\frac{M_*}{M_{*,0}})^{-\gamma}}+\frac{1}{2}
    \end{align} 
    
    where $\log\ M_1=12.58$, $\log\ M_{*,0}=10.9$, $\beta=0.48$, $\delta=0.29$ and $\gamma=1.52$, corresponding to the parameter set in the lowest redshift bin $z=0.1$, along with the scatter of halo mass at fixed stellar mass ($\sigma[M_{\rm h}|M_{\star}]$) values from \citet{Cao2020}, to infer the upper limits of their virial host halo masses. We then estimate their virial radius using the \texttt{Colossus} (\citealp{diemer2018colossus})\footnote{\url{https://bdiemer.bitbucket.io/colossus/toc.html}} cosmology tool. We indicate each candidate's virial radius using a circle in the Figure \ref{fig:environment}. Ultimately, we found no candidate that could convincingly be the host of Zangetsu. 

    Therefore, Zangetsu is likely an isolated, quiescent UDG, unique and rare. While the estimations of halo masses and sizes could be uncertain, we have already adopted the upper limits from today's SHMR relation and use a very `generous' halo boundary definition. However, while Zangetsu is now outside the viral radius of nearby dark matter halos, we cannot rule out the possibility that it had a past interaction with them or the galaxies within them, which could help explain Zangetsu's asymmetric morphology. We will discuss this further in the next section.

\section{Formation Mechanism}
    \label{sec:formation}

    Based on the above analysis, we conclude that Zangetsu is an LSB galaxy with unique and intriguing characteristics that set it apart from known LSB galaxies and are also challenging to explain by conventional theories: 

    \begin{itemize}

        \item Regardless of the exact distance of Zangetsu, the lower limit set by the SBF analysis already puts it on the most diffuse end of the UDG population. 

        \item Moreover, Zangetsu has an unusually elongated morphology, with a low axis ratio $b/a\sim$0.21 outside of the parameter space rarely considered in previous searches for LSBGs or UDGs. 

        \item On top of that, Zangetsu's morphology shows signs of asymmetry, suggesting that it may have been tidally distorted or in interaction with other systems. 

        \item However, despite the rich spectroscopic and multi-wavelength dataset in the COSMOS field, we could not find a convincing host group or galaxy within the plausible redshift range. This means that Zangetsu could be an isolated UDG outside of the viral radius of nearby galaxies.
        
        \item If Zangetsu is indeed isolated, then its quiescent nature (based on its optical color or the overall SED) makes it different from most of the known isolated UDGs (e.g., \citealt{leisman2017alfalfaUDG}). Meanwhile, compared to the few confirmed cases of isolated \& quiescent dwarf galaxies (e.g., \citealp{casey2023Blobby,sand2022tucanaB, Polzin2021cosmosdw1, Karachentsev2015kks3, Makarov2012kkr25, li2024hedgehog, Sand2024Sculptor}), Zangetsu is either much larger or more diffuse.
        
    \end{itemize}

    Putting all these together, we need a formation mechanism that can be applied to an isolated, quiescent, and highly elongated dwarf galaxy that is also extraordinarily diffuse and shows irregular morphology. Needless to say, this is challenging. 

    Firstly, gravitational interaction might be the most convenient channel to explain Zangetsu's diffuse \& quiescent nature, along with any morphological disturbance. For example, Zangetsu could be a tidally disrupting dwarf galaxy in a galaxy group (e.g., \citealt{conselice2018ultra, carleton2019formation, carleton2021excess, Zemaitis2023tidalUDG}). Or it could be formed from the gravitationally bounded part of a large tidal structure in a recent galaxy merger or interaction event (e.g., \citealp{duc2014tdg, lelli2015tdg}), which is supported in simulations (e.g., \citealt{ploeckinger2018tidal}). However, Zangetsu's morphological characters and isolated environment cannot be ignored. Suppose Zangetsu is being disrupted by a surrounding galaxy or dark matter halo. In that case, it should either present a unique ``S-shaped warp'' in the outer part of the galaxy as in the early stage of the disruption (e.g., \citealt{Zemaitis2023tidalUDG, montes2020tidalUDG, Mowla2017, Johnston2002Sshaped}) or display highly off-centered, diffuse light distribution as in the later stage (e.g., \citealt{Okamoto2024tidal, Watts2024}). But despite the moderate asymmetry, Zangetsu's surface brightness distribution follows a very shallow $n\sim 0.4$ profile, with a clear ``outer boundary'' but no sign of isophotal twist in the outskirts. As we discussed earlier, as a ``tidal structure'', the lack of other LSB tidal features and any sign of ``connection'' with the potential host (e.g., the nearby COSMOS~615877 at $z=0.126$) makes it a puzzling case. Plus, typical large tidal features are either ``pointing toward'' the host galaxy, like a tidal tail or stream (e.g., \citealp{Watts2024}), or on the side of the host \& ``bend toward'' it, like a shell (e.g., \citealp{Bilek2022NGC474shell, Foster2014NGC4651shell}). This is also not the case for Zangetsu.

    Next, assuming Zangetsu is genuinely isolated, its quiescent nature and asymmetric morphology become the challenges for any formation scenario. In hydrodynamic simulations, supernova-driven, episodic gas outflows from starbursts can alter the dark matter gravitational potential, leading to the formation of a UDG in isolation (e.g., \citealt{di2017fieldudgSF, jiang2019fieldudgSF}). Additionally, \citealp{jiang2019fieldudgSF} found that field UDGs in the NIHAO simulation tend to be more prolate, consistent with Zangetsu's elongated morphology. Meanwhile, \citealp{van2022UDGspinhalo} suggests that a high-spin dark matter halo can prevent gas from collapsing, contributing to creating a field UDG. However, most isolated UDGs from these mechanisms are less quiescent than Zangetsu. The quenching process often relies on an external factor, such as tidal interaction, and is inconsistent with Zangetsu's current isolated environment. A merger between dwarf galaxies, or between two UDGs, could be another interesting channel to consider (e.g., \citealt{Fielder2023DwarfMerge, Wright2021}), which could help explain Zangetsu's high ellipticity, moderately disturbed morphology, and the lack of star formation. However, the realistic chance of such a rare event outside a galaxy group or cluster environment casts doubt on the feasibility of this explanation.

    Recently, the discovery of more isolated \& quiescent dwarf galaxies (e.g., \citealt{Makarov2012kkr25, Karachentsev2015kks3, Polzin2021cosmosdw1, sand2022tucanaB, casey2023Blobby, li2024hedgehog}) bring the ``backsplash'' process under the spotlight. When a satellite galaxy passes through the pericenter of a more massive dark matter halo, it could experience a gravitational impulse during the interaction and subsequently get ``ejected'' out of the virial radius of the halo (e.g., \citealt{Mamon2004, Gill2005, Wetzel2014}). The backsplash process can effectively act as an environmental quenching mechanism, helping explain the enhanced quenched satellite population near massive halos or isolated quenched dwarf galaxies. At the sub-halo level, the backsplash is a well-known effect in cosmological simulations, but its significance in shaping galaxies' properties is less explored. Recently, \citet{benavides2021backsplash} proposed that the backsplash process can help explain the isolated \& quiescent dwarf galaxy, including UDGs. During the backsplash, tidal interactions could effectively strip gas from a satellite galaxy, accelerating its quenching. In addition to creating a gas-poor dwarf population outside the host halo, in principle, the tidal interaction can help explain the elongated \& asymmetric morphology. So far, candidates of backsplash galaxies are found near the Local Group (e.g., \citealt{McConnachie2012localdwarf, Teyssier2012}) and among the isolated \& quiescent dwarf population (e.g., \citealt{casey2023Blobby, li2024hedgehog}). Using the TNG50 (e.g., \citealt{Nelson2019, Pillepich2019}) high-resolution hydro-simulations, \citet{Bhattacharyya2025} found that the backsplash population can explain the majority of the quenched dwarf galaxies outside 1.5 Mpc from a more massive halo. In theory, Zangetsu could be a backsplash satellite created in past interactions with one of the candidates we identified. Moreover, in the simulated sample of backsplash UDGs from \citet{benavides2021backsplash}, the average distance between UDGs and their hosts is $2.1\ r_{200}$. Ignoring the slight difference between $r_{200}$ and the virial radius, Figure \ref{fig:environment} demonstrates that Zangetsu is located within the $2.1\ r_{200}$ range of all potential host galaxies. This suggests that all the candidates shown in Figure \ref{fig:environment} could plausibly be Zangetsu’s host under this scenario. However, it is worth noting that the virial radius used here is already estimated based on the upper limits of their host halo masses.

    At the same time, as shown in Figure \ref{fig:L_size}, all the backsplash candidates are much less diffuse and have smaller physical sizes than Zangetsu. Also, the most likely backsplash cases, including Cetus, Tucana, Sculptor B (\citealt{Teyssier2012, Sand2024Sculptor}), ``Blobby'' (\citealt{casey2023Blobby}), and ``Hedgehog'' (\citealt{li2024hedgehog}), do not show clear evidence of asymmetric morphology. So, whether Zangetsu could be explained as a backsplash dwarf still needs to be verified. In \citet{Bhattacharyya2025}, the authors noted that the backsplash population in TNG50 is often dark-matter deficient. While it is highly challenging to confirm this for Zangetsu using today's instruments, it would be an important experiment in the future, as the backsplash dwarf could teach us about the environmental quenching process and the satellite's orbit history.

    On the other hand, Zangetsu's uniqueness may result from a selection effect, in which other highly elongated UDGs are excluded before photometric confirmation. The discovery of Zangetsu forces us to reconsider these criteria. Specifically, the approach for excluding tidal tail contamination should be re-evaluated, and morphology-based cuts should be relaxed to identify more Zangetsu analogs.

\section{Conclusions}
    \label{sec:summary}
    
    This paper presents the discovery and analysis of a unique ultra-diffuse galaxy candidate, Zangetsu, in the COSMOS field using HSC imaging data. Zangetsu is an extremely low surface brightness galaxy with $g$-band central surface brightness of $\sim 26.6$ mag/arcsec$^2$ and a very shallow inner light profile (\ser{} index $n\sim 0.40$). Zangetsu's optical color ($g-i=0.96$) and multi-band SED suggest it is a quiescent system. More intriguingly, Zangetsu displays an unusually elongated shape ($b/a\sim 0.25$), unprecedented among the previously known LSB population. While showing a hint of disturbed morphology, with a slightly asymmetric light distribution, Zangetsu is unlikely to interact with nearby systems, including a $z=0.126$ S0 galaxy, which would make Zangetsu a 47 kpc long structure in a nonphysical configuration. Using available HSC and HST images, we constrained Zangetsu's distance to $D>26$ Mpc using the surface brightness fluctuation method. Within a plausible redshift range of $0.006<z<0.1$, Zangetsu sits above the luminosity--size relation of discovered LSB or UDG galaxies regardless of the actual distance, indicating that Zangetsu is very likely to be an extraordinarily diffuse or large system with some unusual features (see Figure \ref{fig:L_size}). At the same time, Zangetsu's isolated environment makes it even more special. Even with the extensive spectroscopic redshift data in the COSMOS field, including the newly released DESI DR1, we find no convincing host of Zangetsu at least within 1.2$\times R_{\rm Vir}$ of a $\log\ M_*/M_{\odot}>9$ galaxy.

    Combining all these, being an isolated, quiescent, large \& diffuse, and highly elongated galaxy with evidence of a disturbed morphology, Zangetsu presents itself as an outstanding challenge for conventional theories of UDG formation. Commonly invoked internal or external formation channels cannot satisfactorily explain \emph{all} of Zangetsu's observed features. Among them, Zangetsu, being a backsplash dwarf galaxy ``ejected'' from a more massive halo through past interaction, could be a potentially competitive scenario that \emph{could} explain Zangetsu's isolation, quiescent nature, and morphological features simultaneously. However, such a scenario still needs to depend on the specific orbital configuration of Zangetsu when it approaches the pericenter of the previous ``host'' massive halo, which needs to be unique enough to create the largest and/or most diffuse backsplash dwarf galaxy known today. In addition to the backsplash picture, we also explore the possibility of a dwarf-dwarf major merger, an isolated UDG formed in a high-spin halo, and remnant extreme tidal interaction scenarios. Yet, none of them match Zangetsu well. Follow-up observations, particularly deep spectroscopy, will be crucial to confirm Zangetsu's distance \& physical size and, more importantly, to pinpoint its physical nature and reveal its most likely formation scenario.

    Beyond the serendipitous discovery of a rare isolated quenched UDG, Zangetsu highlights a potential selection bias in today's LSB/UDG hunt: as shown in Figure \ref{fig:morp_comp}, the highly elongated population ($b/a<0.4$) might be extremely rare, especially when they are outnumbered by contamination from tidal features and edge-on disc galaxies, but they could help us extend our understanding of the formation \& evolution of the LSB galaxies or even the dwarf galaxies in general. 
    

\section{Acknowledgment}

  SH acknowledges support from the National Natural Science Foundation of China Grant No. 12273015 \& No. 12433003 and the China Crewed Space Program through its Space Application System. 

  The authors thank Annika Peters, Joy Bhattacharyya, Pieter van Dokkum, and Aaron Romanowsky for useful discussions and comments.
  
  The Hyper Suprime-Cam (HSC) collaboration includes the astronomical communities of Japan, Taiwan, and Princeton University.  
  The HSC instrumentation and software were developed by the National Astronomical Observatory of Japan (NAOJ), the Kavli Institute for the Physics and Mathematics of the Universe (Kavli IPMU), the University of Tokyo, the High Energy Accelerator Research Organization (KEK), the Academia Sinica Institute for Astronomy and Astrophysics in Taiwan (ASIAA), and Princeton University. Funding was contributed by the FIRST program from Japanese Cabinet Office, the Ministry of Education, Culture, Sports, Science and Technology (MEXT), the Japan Society for the Promotion of Science (JSPS), Japan Science and Technology Agency (JST), the Toray Science Foundation, NAOJ, Kavli IPMU, KEK, ASIAA, and Princeton University.
  
  This research used data obtained with the Dark Energy Spectroscopic Instrument (DESI). DESI construction and operations are managed by the Lawrence Berkeley National Laboratory. This material is based upon work supported by the U.S. Department of Energy, Office of Science, Office of High-Energy Physics, under Contract No. DE–AC02–05CH11231, and by the National Energy Research Scientific Computing Center, a DOE Office of Science User Facility under the same contract. Additional support for DESI was provided by the U.S. National Science Foundation (NSF), Division of Astronomical Sciences under Contract No. AST-0950945 to the NSF’s National Optical-Infrared Astronomy Research Laboratory; the Science and Technology Facilities Council of the United Kingdom; the Gordon and Betty Moore Foundation; the Heising-Simons Foundation; the French Alternative Energies and Atomic Energy Commission (CEA); the National Council of Science and Technology of Mexico (CONACYT); the Ministry of Science and Innovation of Spain (MICINN), and by the DESI Member Institutions: www.desi.lbl.gov/collaborating-institutions. The DESI collaboration is honored to be permitted to conduct scientific research on Iolkam Du’ag (Kitt Peak), a mountain with particular significance to the Tohono O’odham Nation. Any opinions, findings, and conclusions or recommendations expressed in this material are those of the author(s) and do not necessarily reflect the views of the U.S. National Science Foundation, the U.S. Department of Energy, or any of the listed funding agencies.
  
  This work has made use of data from the European Space Agency (ESA) mission {\it Gaia} (\url{https://www.cosmos.esa.int/gaia}), processed by the {\it Gaia} Data Processing and Analysis Consortium (DPAC, \url{https://www.cosmos.esa.int/web/gaia/dpac/consortium}). Funding for the DPAC has been provided by national institutions, in particular the institutions participating in the {\it Gaia} Multilateral Agreement.

  Some of the data presented in this paper were obtained from the Mikulski Archive for Space Telescopes (MAST) at the Space Telescope Science Institute. The specific observations analyzed can be accessed via \dataset[https://doi.org/10.17909/n5ad-3996]{https://doi.org/10.17909/n5ad-3996}. STScI is operated by the Association of Universities for Research in Astronomy, Inc., under NASA contract NAS5–26555. Support to MAST for these data is provided by the NASA Office of Space Science via grant NAG5–7584 and by other grants and contracts.
 
  This research made use of:
  
  \href{http://www.scipy.org/}{\texttt{SciPy}},
      an open-source scientific tool for Python (\citealt{SciPy});
  \href{http://www.numpy.org/}{\texttt{NumPy}}, 
      a fundamental package for scientific computing with Python (\citealt{NumPy});
  \href{http://matplotlib.org/}{\texttt{Matplotlib}}, 
      a 2-D plotting library for Python (\citealt{Matplotlib});
  \href{http://www.astropy.org/}{\texttt{Astropy}},
      a community-developed core Python package for Astronomy (\citealt{astropy:2013, astropy:2018, astropy:2022}); 
  \href{https://ipython.org}{\texttt{IPython}}, 
      an interactive computing system for Python (\citealt{IPython});
  \href{https://photutils.readthedocs.io/en/stable/index.html}{\texttt{Photutils}},
      an Astropy package for detection and photometry of astronomical sources(\citealt{photutils}); 
  \href{http://bdiemer.bitbucket.org/}{\texttt{Colossus}}, 
      COsmology, haLO and large-Scale StrUcture toolS (\citealt{Colossus});
  \href{https://drizzlepac.readthedocs.io/en/latest}{\texttt{AstroDrizzle}} \citep{Drizzle2002, Drizzle2010}.
  \href{https://astrophot.readthedocs.io/en/latest/index.html}{\texttt{AstroPhot}}\citep{astrophot}


\bibliography{zangetsu}


\appendix
\section{The multi-band images of Zangetsu}
\label{app:multiband}

\begin{figure*}[t!]
    \centering 
    \includegraphics[width=\textwidth]{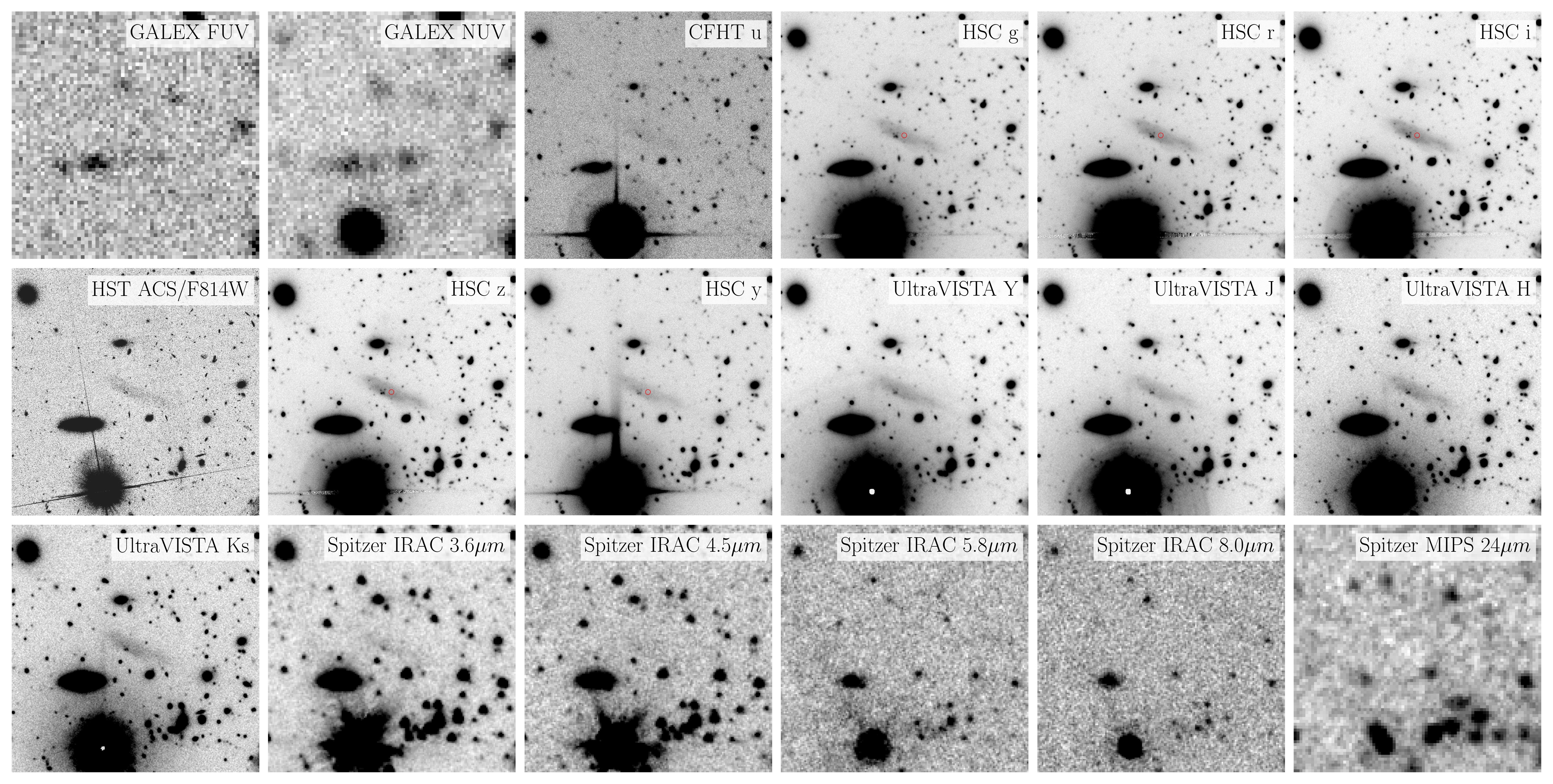}
    \caption{Cutouts of Zangetsu in selected filter band from ultraviolet to mid-infrared, including GALEX, CFHT, HSC, HST, UltraVISTA, and Spitzer detections. The red circle in the center of the HSC image is the $2\arcsec$ aperture used in the COSMOS2020 catalog (\citealp{weaver2022cosmos2020}).
    }
        
    \label{fig:multiband_image}
\end{figure*} 

    In Figure \ref{fig:multiband_image}, we summarize the most relevant public multi-wavelength images of Zangetsu across a series of representative filters from the ultraviolet to the mid-infrared, including detections from GALEX, CFHT, HSC, HST, UltraVISTA, and Spitzer. Each cutout is centered on the position of Zangetsu, with a size of $100\arcsec\times 100\arcsec$. In addition to the broadband images we show here, Zangetsu has also been detected in 12 intermediate bands and 2 narrow bands from Subaru Suprime-Cam, which are omitted here because they do not provide additional information about Zangetsu, given its quiescent nature.

\section{Systematic Impact of the 1-D Surface Brightness Profile}
\label{app:sky}

\begin{figure}[t!]
    \centering
    \begin{minipage}[t]{0.9\linewidth}  
        \centering
        \label{fig:iso_star_g}\includegraphics[width=6in]{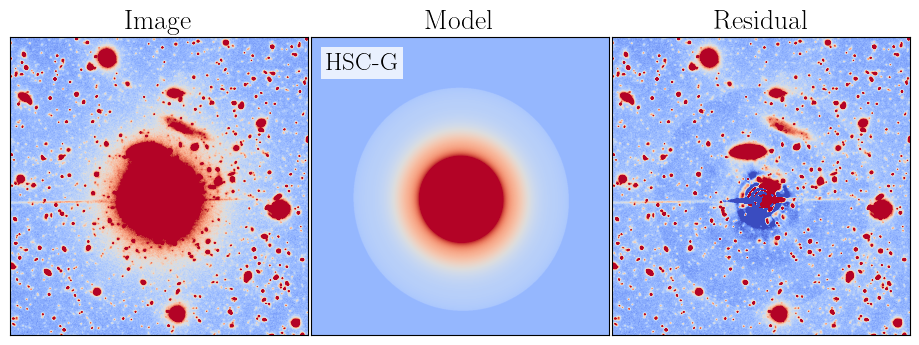}

    \end{minipage}

    \begin{minipage}[t]{0.9\linewidth}
        \centering
        \label{fig:iso_star_r}\includegraphics[width=6in]{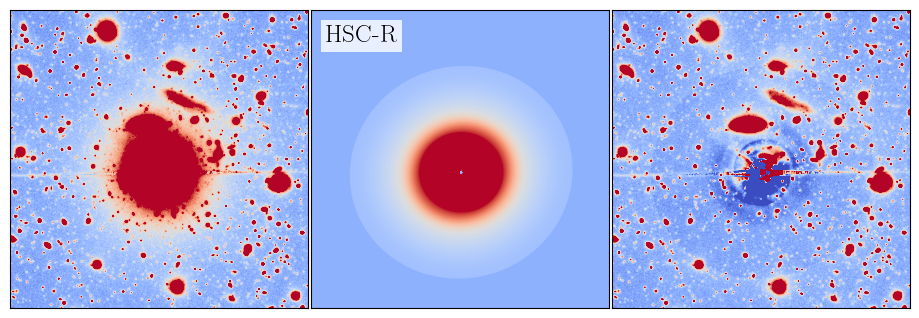}
    \end{minipage}

    \begin{minipage}[t]{0.9\linewidth}
        \centering
        \label{fig:iso_star_i}\includegraphics[width=6in]{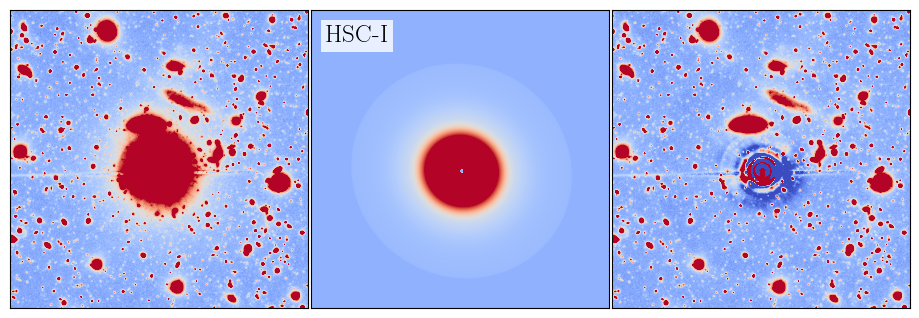}
    \end{minipage}
    \caption{
        We illustrate the subtraction of scattered light around the bright star in the $gri$ bands. The left panels present the original images, centered on the star and sized $168\arcsec \times 168\arcsec$. The middle panels show the 2-D model reconstructed from the 1-D isophote fitting. The small ``holes'' in the $r$- and $i$-band models are due to a known numerical issue when reconstructing the 2-D model based on the 1-D surface brightness profile using \texttt{photutils.isophote}. This does not affect the subtraction of scattered light or Zangetsu's modeling. The right panels are the residual map after subtracting the 2-D model. The central residual exhibits complex, asymmetric features resulting from the stacking of multiple images. However, these residual patterns do not affect the correction of the large-scale background level around Zangetsu.
    }
    \label{fig:star_resdiual}
  \end{figure}

  \begin{figure*}[htbp]
    \centering 
    \includegraphics[width=0.7\textwidth]{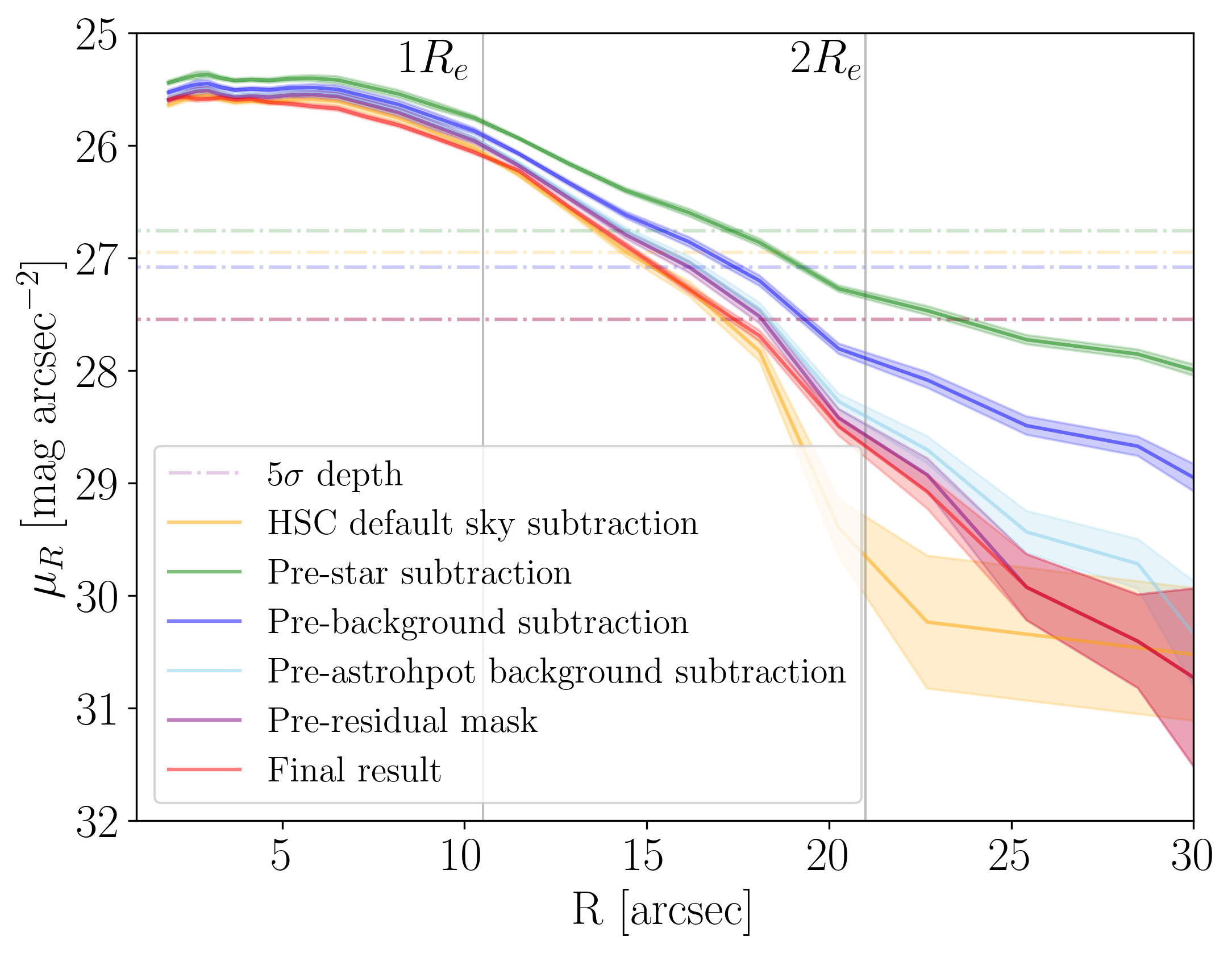}
    \caption{
        Here, we illustrate the impacts of different imaging systematics on the surface brightness profile of Zangetsu. These profiles also reflect the result of each correction step adopted by this work. We highlight the 5-$\sigma$ surface brightness limits in each stage using dot-dashed lines with the corresponding colors. The vertical lines indicate $1 \times$ and $2 \times$ effective radius $R_e$ in the final result. 
    }     
    \label{fig:1D_profile_test}
\end{figure*}

    The LSB nature of Zangetsu and its complex surroundings make it challenging to obtain a reliable surface brightness profile, as different imaging systematics and their correction procedures can significantly affect the results. As explained in Sec. \ref{sec:photometry}, we have taken special care of the contamination from a nearby saturated star, the local background level, and the masking of nearby objects. Here, we briefly summarize each step adopted in preparing the final image for surface brightness modeling and illustrate their impact in Figure \ref{fig:1D_profile_test}:
        
\begin{itemize}
    \item For the coadded HSC images, we adopted the version based on full focal plane background modeling (or ``global'' sky subtraction) to avoid the significantly over-subtracted background commonly seen in the local sky subtracted version, which is the default option in HSC data cutout. However, this does not guarantee that the background around Zangetsu is not over- or under-subtracted.
    
    \item We subtract a model of the nearby bright star Gaia DR3 3834532164620977152 with $m_G=12.04$ mag to reduce the impact of its extended scattered light on the surface brightness profile recovery of Zangetsu. The star is $\sim 40\arcsec$ away from Zangetsu. We use the \textit{photutils.isophote} function to model the surface brightness distribution of the star and subtract it from the image. While this isophote-based model can help remove the extended scattered light, it cannot perfectly describe the detailed light distribution of a saturated star. Therefore, we still mask out the star for the following modeling process.
    
    \item We then estimate the local sky background using the \textit{MedianBackground} algorithm from \textit{Photutils} and correct the image to ensure the background region surrounding Zangetsu has no over- or under-subtraction. We adopt a $15\times15$ box size and perform $3\sigma$ sigma clipping. Adjusting these parameters within a reasonable range will not significantly affect the main results.
    
    \item Next, we independently perform initial 2-D fitting to Zangetsu in each band. Based on these results, we aggressively mask nearby objects from the residual maps. The combination of the masks from each band and the bright star mask is used in the final 2-D fitting. 

    \item Finally, using \texttt{AstroPhot}, we jointly fit a \ser{} model and a sky background model (tilted plane) to the $gri$ band images. In addition to the \ser{} parameters that describe Zangetsu's 2-D light distributions, the result also provides the final ``global'' background estimation around it. After correcting this 2-D sky background model from the image, we extract the 1-D profile of Zangetsu based on the geometric parameters from the \ser{} model (centroid, position angle, and axis ratio).

\end{itemize}

    To demonstrate the exact effect of the bright-star subtraction, we show the original images, the reconstructed 2-D models from isophote fitting, and the residual maps in the $gri$ bands in Figure \ref{fig:star_resdiual}. We extend the 1-D profile model to $\sim67\arcsec$ from the center to ensure the removal of most of the scattered light from the star on Zangetsu. 

    To evaluate the impact of each step described above on the final 1-D profile, we extract the 1-D profiles at different stages of background subtraction and masking using the morphological parameters derived from the final 2-D \ser{} model. We present the comparisons of the 1-D profiles in Figure \ref{fig:1D_profile_test}, clearly showing the necessity and importance of these correction steps. Without correcting for the global background and removing scattered light, we will significantly overestimate the surface brightness and size of Zangetsu's outskirts. The masking of nearby objects and the final correction of the 2-D sky background model from \texttt{AstroPhot} have limited impact on the final surface brightness profile. Meanwhile, the coadded HSC image with aggressive local sky subtraction results in significant over-subtraction beyond $1.5 \times R_e$ in Zangetsu by 0.5 to 1 mag/arcsec$^2$. 

    We should note that these correction steps are empirical. However, it is worth noting that, if our background correction or scattered light removal steps were too aggressive, Zangetsu would have a slightly increased surface brightness level within $R_e$ and total luminosity, but with an \emph{even larger effective radius}, as these corrections are more critical to the low surface brightness outskirts of Zangetsu. Nevertheless, none of these systematics will change the conclusion that Zangetsu is an intriguing outlier on the luminosity-size relation regardless of distance.

    In addition, we estimate the surface brightness limits by averaging the RMS map values before each processing step. These limits are plotted as horizontal dash-dotted lines in Figure \ref{fig:1D_profile_test}, demonstrating that each step significantly improves the surface brightness limit. The final 5-$\sigma$ surface brightness limitation reaches approximately $\sim 27.54$ mag/arcsec$^2$ in $g$-band within a $\sim 2.52\arcsec$ angular scale.

\section{Injection Recovery Test of Zangetsu}
\label{app:injection}
\begin{figure*}[htbp]
    \centering 
    \includegraphics[width=\textwidth]{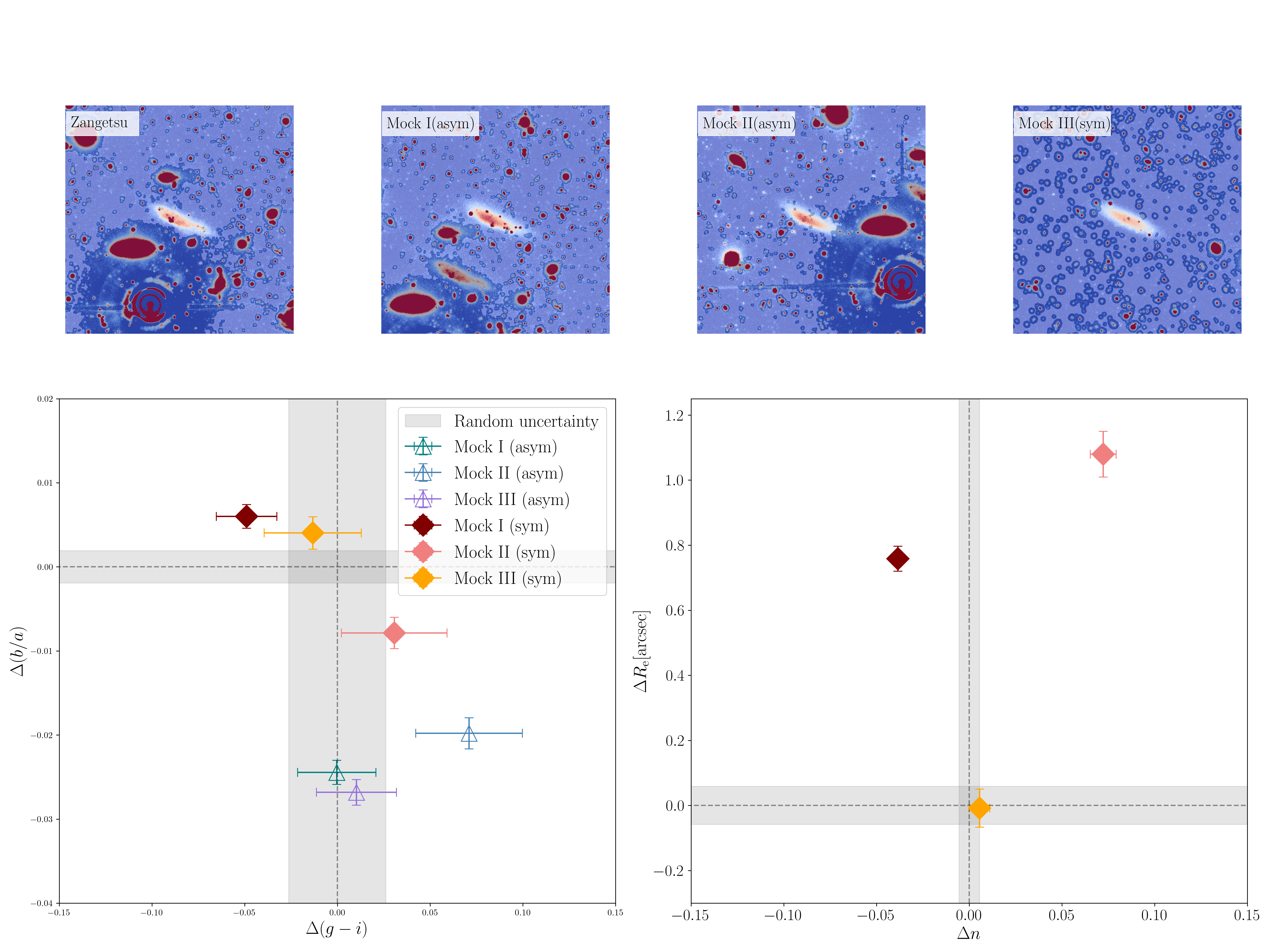}
    \caption{\textbf{Upper panel:} Images of Zangetsu and the mock-injected galaxies at three positions. The models shown in Mock I and Mock II are asymmetric, whereas the model in Mock III is symmetric. 
    \textbf{Lower panel:} Differences between the fitted and input parameters. The gray-shaded region indicates the statistical errors derived from the best-fitting model.}
     
    \label{fig:mock_test}
\end{figure*}

    To evaluate potential systematics in our fitting results, we perform an injection-recovery test for Zangetsu. We select three positions for model injection to test the possible effects of the varying sky background caused by the nearby bright star. The first two positions are placed on the scattered light removed image: one at the same angular distance from the bright star but on the opposite side, and the other along the same orientation but at a larger angular distance. We also chose a nearby clean sky region sufficiently far from the bright star as the third position to avoid any background-related impact.

    We inject two models into each of these positions. Given the asymmetric residual features shown in Figure~\ref{fig:profile}, we first fit Zangetsu with a \ser{} model with Fourier modes using \texttt{Galfit} to derive the input parameters. We also inject a symmetric model based on Zangetsu’s best-fitting parameters listed in Table~\ref{tab:param}. Each mock object is then processed through the same pipeline as Zangetsu, including source de-blending, local sky subtraction, and MCMC fitting with a single \ser{} model.

    The resulting mock images and the differences between the fitted and input parameters are shown in Figure~\ref{fig:mock_test}. We compare key parameters color $g-i$, axis ratios $b/a$, \ser{} index, and effective radius $\rm{R}_e$, which are essential for our conclusion that Zangetsu is a quiescent and extremely elongated galaxy. Because the definition of effective radius differs between a single \ser{} model and a \ser{} model with Fourier models, the asymmetric model injection results are not plotted in the right panel.

    This test demonstrates that uncertainties introduced by the local sky background dominate over statistical fitting errors. Taking the largest difference as our estimate, the maximum relative errors from the mock tests are $7.4\%$ for $(g-i)$, $10.7\%$ for $(b/a)$, $18.1\%$ for $n$, and $10.3\%$ for $\rm{R}_{e}$. Nevertheless, the key conclusion remains robust: Zangetsu exhibits an unusually elongated morphology for a UDG, red color indicating a quiescent old stellar population, and a relatively large physical size for its redshift. The differences caused by the sky background in Figure~\ref{fig:mock_test} do not hurt this conclusion. 

\section{Posterior distribution of the \ser{} model}
\label{app:corner}

  \begin{figure*}[t!]
    \centering 
    \includegraphics[width=\textwidth]{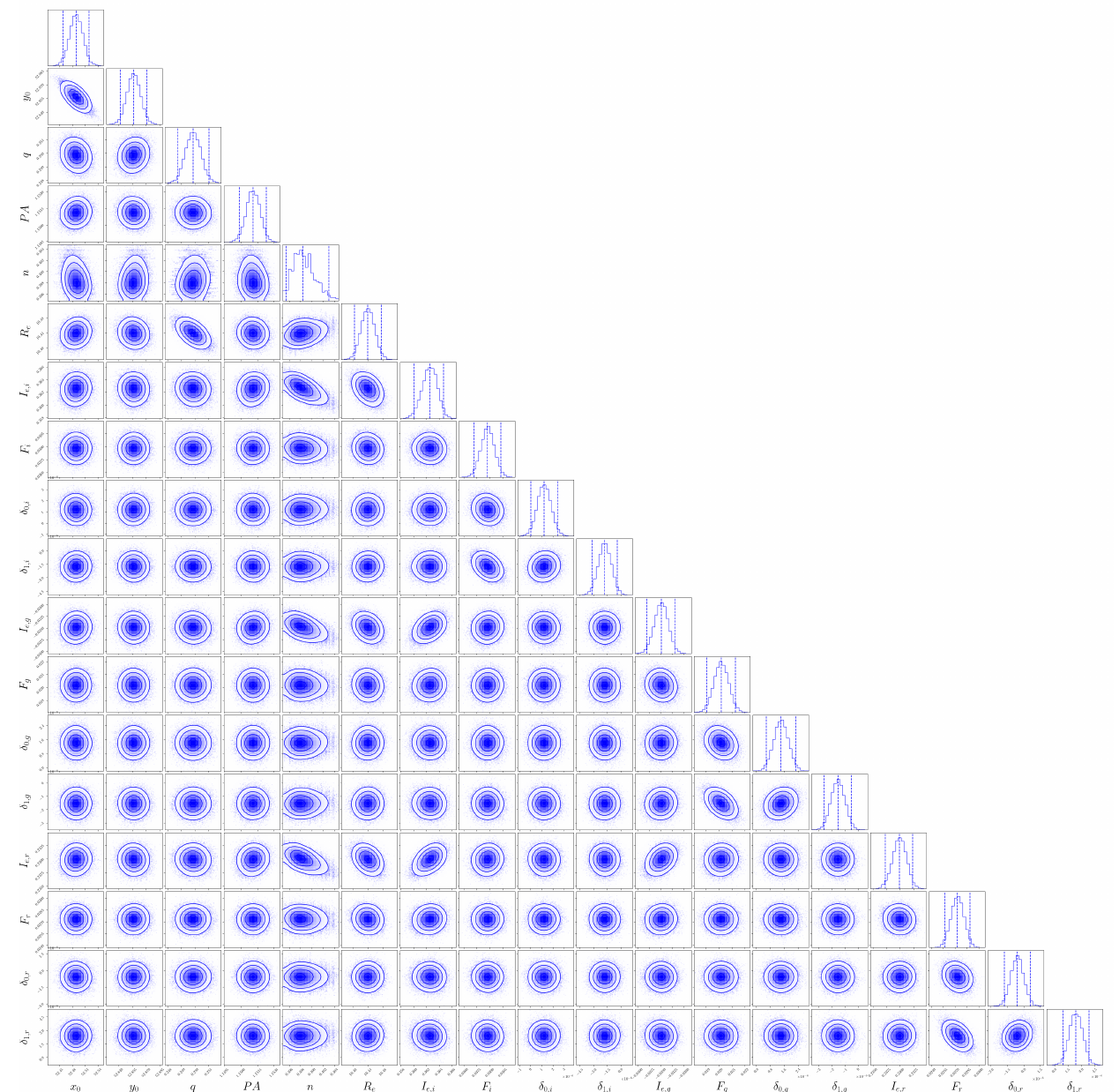}
    \caption{The posterior distribution from the \ser{} + plane sky model of Zangetsu.}     
    \label{fig:corner}
\end{figure*}

Figure \ref{fig:corner} presents the posterior distribution obtained from the MCMC sampling described in Sec. \ref{ssec:galaxy_model}. The model is composed of a joint \ser{ } model across the $gri$ bands, combined with the corresponding plane sky background model. In total, 18 parameters are fitted:

\begin{itemize}
    \item \ser{} model parameters: central position ($x_0$, $y_0$), effective radius $R_e$, \ser{} index $n$, axis ratio $b/a$ and the surface brightness at the effective radius $I_{e,x}$ for each band.
    \item Sky background parameters: for each band, the sky background is modeled with a plane sky model defined by a central flux level $F_x$ and two gradient terms $\delta_{0,x}$ and $\delta_{1,x}$ representing the slopes along the two spatial axes. 
    
\section{SED fitting of Zangetsu using Mephisto}
    \label{app:sed}
    
\end{itemize}
    While the exact distance to Zangetsu remains unknown, the upper limit on the distance and the possible redshift range can already support a preliminary SED fit. Assuming Zangetsu is fixed at $z=0.1$, the upper limit of the redshift range we estimated in Sec. \ref{sec:results}, we compile the multiband SED of Zangetsu from the \texttt{CLASSIC} COSMOS2020 catalog (\citealp{weaver2022cosmos2020}) based on the PSF-matched aperture photometry. We then fit the SED using the \texttt{CIGALE} (\citealp{Boquien2019cigale}) code with the help of a large language model-based agent \texttt{Mephisto} (\citealp{Sun2024mephisto}). Compared to manual fitting, \texttt{Mephisto} can automatically explore different modeling options and prior choices for \texttt{CIGALE}. Through an iterative process of knowledge distillation and validation, \texttt{Mephisto} can simultaneously select the most promising model and test the robustness of the fit. 

\begin{figure*}[htbp]
    \centering
    \begin{minipage}[b]{0.48\textwidth}
        \includegraphics[width=\textwidth]{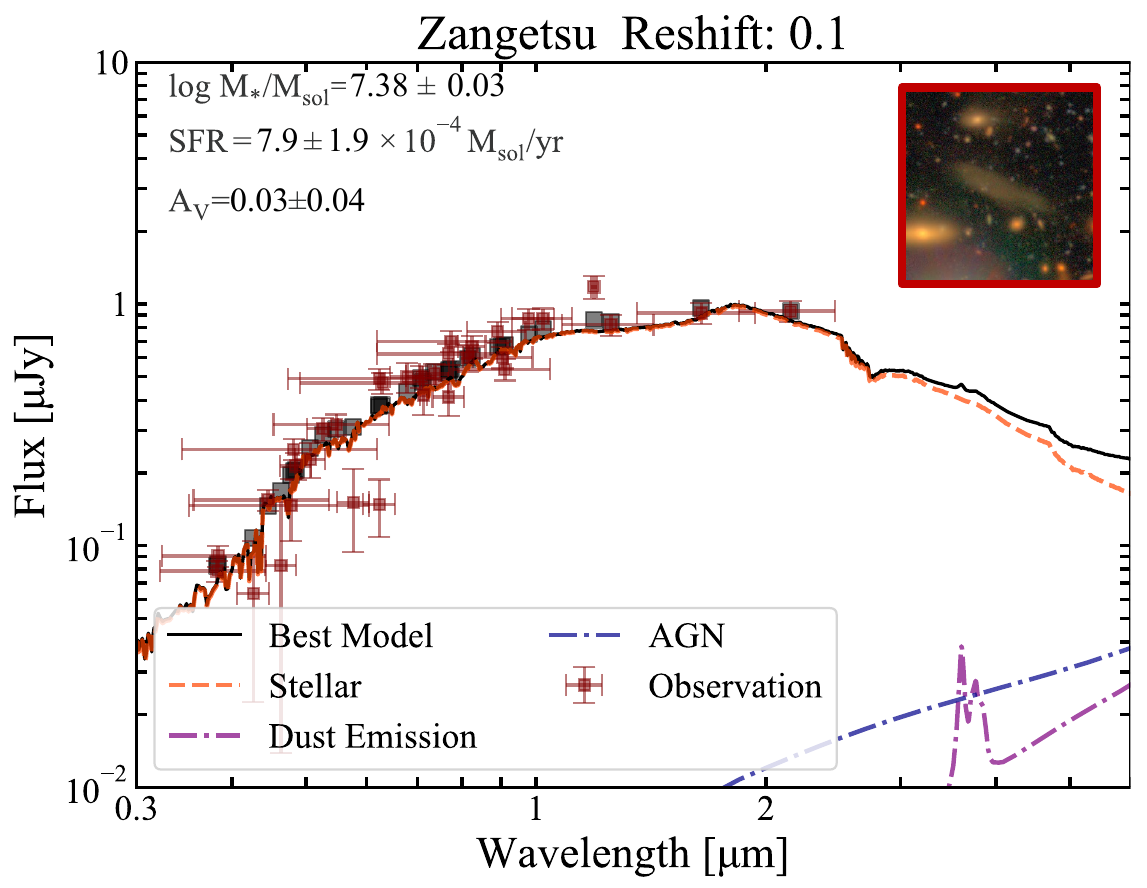}
        \label{fig:zangetsu_sed}
    \end{minipage}
    \hfill
    \begin{minipage}[b]{0.48\textwidth}
        \includegraphics[width=\textwidth]{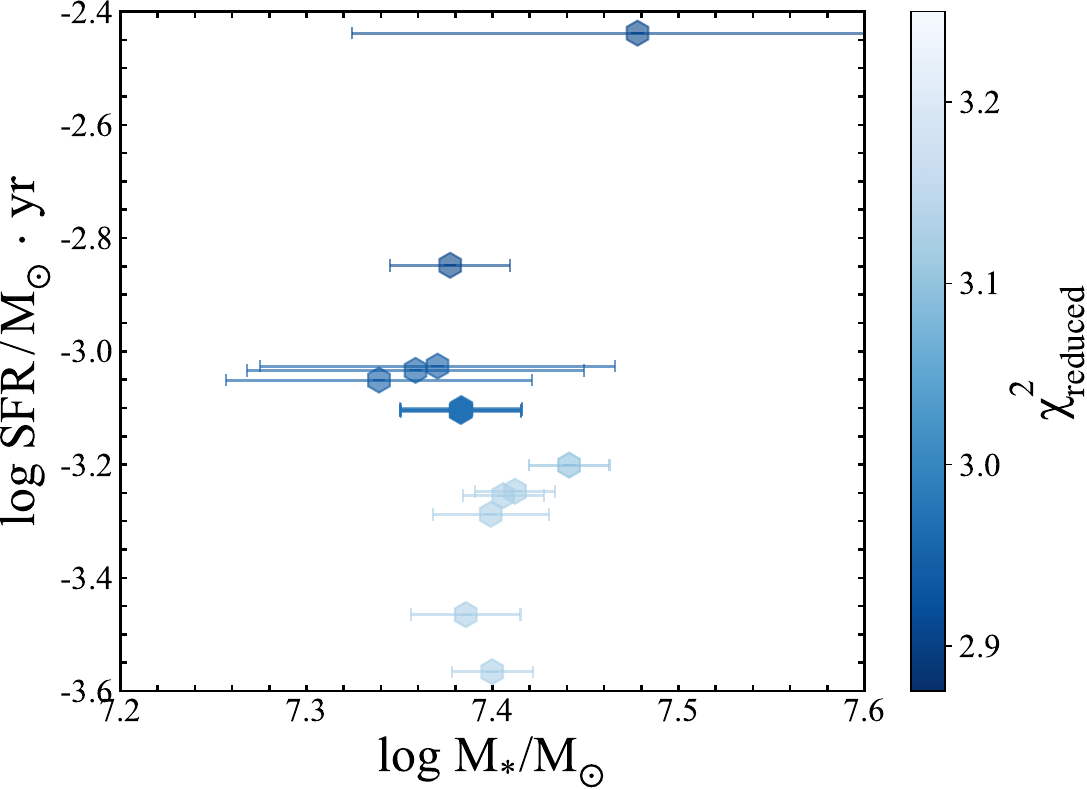}
        \label{fig:zangetsu_solution}
    \end{minipage}
    
    \caption{\textbf{Left Panel:} The SED fitting result of Zangetsu using Mephisto. \textbf{Right Panel:} Various posterior SED solutions sampled by Mephisto in the stellar mass and star formation rate parameter space. All solutions shown here reflect Zangetsu's quenched nature. Note that error bars are shown on the y-axis, but they are barely noticeable due to their small size.}
    
    \label{fig:sed_res}
\end{figure*}

    Figure \ref{fig:sed_res} displays Zangetsu's SED and the best-fit model from \texttt{Mephisto}. On the right side of Figure \ref{fig:sed_res}, we also show \texttt{Mephisto}'s summary of all the promising models explored on the stellar mass-star formation rate plane, color-coded by the reduced $\chi^2$ of the model, which confirms that Zangetsu is quiescent. For the best-fit model, \texttt{Mephisto} suggests that, under the assumption of a Chabrier IMF (\citealp{Chabrier2003IMF}) and delayed-$\tau$ star formation history model, Zangetsu’s SED can be represented by an age of 9 Gyr old stellar population that experienced an early, rapid episode of star formation followed by a long quiescent phase, with extremely low SFR ($<1\times 10^{-3}\ M_{\odot}/{\rm yr}$) and dust extinction ($A_{V}\sim 0$).

    Using the $M/L$ estimated based on the SED fitting and the $i$-band magnitude of the 2-D \ser{} model, Zangetsu should have a stellar mass in the range of 6.91 $<\log M_\star/M_{\odot}<$ 7.64 at $z=0.1$, and 8.57 $<\log M_\star/M_{\odot}<$ 9.30 after aperture correction, broadly consistent with the predictions based on optical color after aperture correction. 

\end{CJK*}


\label{lastpage}

\end{document}